\PassOptionsToPackage{table,svgnames, x11names}{xcolor} 
\PassOptionsToPackage{scaled=.8}{beramono} 
\documentclass[a4paper,USenglish,cleveref, autoref, thm-restate]{lipics-v2021} 

\usepackage[T1]{fontenc}
\usepackage[utf8]{inputenc}

\usepackage[scaled=.8]{beramono} 
\usepackage[noteubner, frozen]{showcode}

\usepackage[table,svgnames,dvipsnames]{xcolor}
\usepackage{pifont}

\newcommand{\dinggreen}[1]{{\color{ForestGreen}\ding{#1}}}
\newcommand{\dingblue}[1]{{\color{Blue}\ding{#1}}}

\definecolor{MyRawSienna}{RGB}{173, 52, 50}
\definecolor{MyOrange}{RGB}{254, 113, 28}
\definecolor{MyBloodRed}{RGB}{153, 0, 100}
\definecolor{MyTealBlue}{RGB}{1, 172, 175}
\definecolor{MyRoyalBlue}{RGB}{1, 87, 180}
\definecolor{MyViolet}{RGB}{61, 60, 158}
\definecolor{NegativeMyViolet}{RGB}{158, 60, 61}

\definecolor{Fig3Blue}{RGB}{70, 143, 227}
\definecolor{Fig3Orange}{RGB}{217, 114, 73}
\usepackage{paralist}
\usepackage{footmisc}
\usepackage{afterpage}
\usepackage{algorithm}
\usepackage{algpseudocode}
\newcommand{\ourtool}[1]{\textsc{#1}}

\usepackage{tikz}

\usepackage{amsmath}
\usepackage{pdflscape}

\usepackage{soul}
\newcommand{\hlc}[2][yellow]{\sethlcolor{#1}\hl{\,#2\,}}

\usepackage{booktabs}

\title{Prioritizing Configuration Relevance via Compiler-Based Refined Feature Ranking}

\author{Federico {Bruzzone}}{Universit\`a degli Studi di Milano, Milan, Italy}{federico.bruzzone@unimi.it}{0009-0004-6086-8810}{}
\author{Walter {Cazzola}}{Universit\`a degli Studi di Milano, Milan, Italy}{cazzola@di.unimi.it}{0000-0002-4652-8113}{}
\author{Luca {Favini}}{Universit\`a degli Studi di Milano, Milan, Italy}{luca.favini1@studenti.unimi.it}{}{}
\authorrunning{F. Bruzzone, W. Cazzola, L. Favini}
\Copyright{Federico Bruzzone, Walter Cazzola, and Luca Favini}



\category{} 
\relatedversion{} 

\supplement{A replication package is available on Zenodo.}
\supplementdetails[]{Software}{https://doi.org/10.5281/zenodo.17691776}

\funding{This work was partially supported by the MUR project ``T-LADIES'' (PRIN 2020TL3X8X).} 

\nolinenumbers 

\EventEditors{John Q. Open and Joan R. Access}
\EventNoEds{2}
\EventLongTitle{42nd Conference on Very Important Topics (CVIT 2016)}
\EventShortTitle{CVIT 2016}
\EventAcronym{CVIT}
\EventYear{2016}
\EventDate{December 24--27, 2016}
\EventLocation{Little Whinging, United Kingdom}
\EventLogo{}
\SeriesVolume{42}
\ArticleNo{23}

\begin{document}
\maketitle

\begin{abstract}
Modern programming languages, most notably Rust, offer advanced linguistic constructs for building highly configurable software systems as aggregation of features---identified by a configuration. However, they pose substantial challenges for program analysis, optimization, and testing, as the combinatorial explosion of configurations often makes exhaustive exploration infeasible. In this manuscript, we present the first compiler-based method for prioritizing configurations. Our approach consists of four main steps: \begin{inparaenum}
    \item extracting a tailored intermediate representation from the Rust compiler,
    \item constructing two complementary graph-based data structures,
    \item using centrality measures to rank features, and
    \item refining the ranking by considering the extent of code they impact.
\end{inparaenum}
A fixed number of most relevant configurations are generated based on the achieved feature ranking.
The validity of the generated configurations is guaranteed by using a SAT solver that takes a representation of this graph in conjunctive normal form.
We formalized this approach and implemented it in a prototype, \ourtool{RustyEx}, by instrumenting the Rust compiler. An empirical evaluation on higher-ranked open source Rust projects shows that \ourtool{RustyEx} efficiently generates user-specified sets of configurations within bounded resources, while ensuring soundness by construction. The results demonstrate that centrality-guided configuration prioritization enables effective and practical exploration of large configuration spaces, paving the way for future research in configuration-aware analysis and optimization.
\end{abstract}

\section{Introduction}\label{sect:intro}

\smallskip\noindent\textbf{Premise.}\quad
Highly configurable software systems\footnote{%
   These systems are also known as \textit{product families} or \textit{software product lines} (SPLs)~\cite{Clements01}
} often aim to satisfy a wide range of requirements.
These systems provide a set of features that can be combined in different ways to create a variety of products~\cite{Apel13b}.
Variability-rich software system development leverages principles from product line engineering, commonly referred to as \textit{feature-oriented programming}~\cite{Prehofer97}.
Programming languages, such as C/C++ and Java, provide mechanisms to manage software variability via preprocessor directives, conditional compilation, and annotations. Rust~\cite{Matsakis14} has also been designed to support the development of highly configurable software systems via its \textit{attribute system}\footnote{%
   Rust's attributes are a form of syntactic metadata that can be attached to various parts of the code. Two kind of attributes are supported: \textit{outer} and \textit{inner}. The former is used to annotate \textit{item} declarations (such as functions, structs, and enums), \textit{expression} and \textit{statement}. The latter is used to annotate \textit{modules} and \textit{block expressions} (in particular cases). We will refer to the \textit{items}, \textit{expressions}, and \textit{statements} as \textit{terms}. See \url{https://doc.rust-lang.org/reference} for more details.%
}---specifically, the \inlinerust{cfg} attribute.

\smallskip\noindent\textbf{Problem Statement.}\quad
Software variability can be complex, making it challenging to reason about all possible configurations~\cite{Agh24}.
This problem is further exacerbated by the fact that the number of possible configurations grows exponentially with the number of features, as demonstrated by Krueger~\cite{Krueger06}.
For instance, the Linux kernel is a well-known highly configurable software system with a significant number of features~\cite{Sincero07} that exploits the native support for variability in C/C++ and, more recently, Rust. Over the years, the kernel and similar systems have been extensively studied, with numerous works highlighting the challenges posed by large configuration spaces~\cite{Melo16, Tartler11, ElSharkawy17, Martin21}.

\smallskip\noindent\textbf{State of the Art}\quad
Exhaustive approaches become infeasible due to the combinatorial explosion of configurations~\cite{Apel11}. Therefore, a strategy is needed to reduce the number of configurations under analysis. Several approaches aim to achieve this goal. Sampling techniques~\cite{Patel13, AlHajjaji19, Lee19}, combinatorial testing~\cite{Oster10, Lochau12}, and in particular t-wise testing~\cite{Cohen07, Perrouin10, Oster10} attempt to reduce the configuration space while ensuring coverage. Other works have proposed prioritization methods~\cite{ElSharkawy19, Sanchez14, Parejo16, ElSharkawy17}, using criteria such as similarity~\cite{AlHajjaji19}, non-functional properties~\cite{Sanchez17}, or centrality-based measures~\cite{Mohammed24, Peng16, Levasseur24, Bagheri10}. As reported by Classen \textit{et al.}~\cite{Classen13}, many configurable systems are safety-critical, which makes prioritization even more important: the goal is to detect faults as early as possible while reducing the number of tests.

\smallskip\noindent\textbf{Limitations of Existing Approaches.}\quad
Despite these efforts, the problem of finding a proper prioritization criterion remains open. Existing approaches \begin{inparaenum}
   \item apply only to feature models~\cite{Benavides10, Pohl11}, without considering dependencies in the code, and
   \item do not account for the extent of the code affected by each feature.
\end{inparaenum}
Consequently, configurations that are critical to the system’s behavior may be omitted.
Moreover, the need for principled configuration prioritization extends well beyond testing. Highly configurable systems must also address:
\begin{compactenum}
   \item \textit{compiler optimizations} and \textit{performance analysis}, where either only a subset of variants can be feasibly evaluated, or configuration relevance can guide optimization strategies~\cite{Velez20, Siegmund15, Tartler14, Velez21},
   \item \textit{program comprehension} and \textit{debugging}, where developers must reason about representative or critical configurations~\cite{Halin19, Velez22},
   \item \textit{variability management} and \textit{refactoring}, where structural changes should be guided by the actual impact of features in the code~\cite{Liebig15, Liebig17}, and
   \item \textit{regression analysis}, where prioritization helps identify which configurations are most likely to reveal behavioral differences after evolution~\cite{Qu08, Souto18}.
\end{compactenum}
In all these contexts, treating features uniformly or relying on stochastic heuristics is insufficient.

\smallskip\noindent\textbf{Proposal.}\quad
In this paper, we propose---to the best of our knowledge---the first general method for prioritizing the relevance of configurations via a compiler-based refined ranking of features.
To this end, we present \ourtool{RustyEx}, a fully automated tool designed to identify the \textit{most relevant configurations} in highly configurable Rust software systems.\footnote{%
   Although we use Rust in this work, our approach to configuration prioritization is language-agnostic and applicable to any language with native variability support, such as C/C++ or Java.
}
We define these configurations as those containing the most relevant features. A feature's relevance is measured by its centrality in the \textit{feature dependency graph} and the extent of the code it impacts.
As shown in Figure~\ref{fig:architecture}, we instrument the Rust compiler by performing an interprocedural static analysis~\cite{Olender90, Olender92}\footnote{Inter-procedural analysis is a static analysis technique that analyzes the control and data flow of a program across multiple functions.} to extract the \textit{unified intermediate representation} (UIR).
The UIR is obtained by removing irrelevant information from the Rust \textit{abstract syntax tree} (AST) and by creating the \textit{atoms}.\footnote{%
    From now on, we refer to an \textit{atom} as a single \textit{term} annotated with a given \textit{configuration predicate}.
}
\ourtool{RustyEx} performs static analysis on the UIR to extract two main data structures\@: the \textit{dependency graph}~\cite{Mazurkiewicz95} and the \textit{polytree}~\cite{Dasgupta99}.
The former is a weighted directed graph, dubbed \textit{feature dependency graph} (see \dinggreen{183} and \dingblue{183} in Figure~\ref{fig:rust_and_graphs}), that represents the dependencies between the features.
The latter is an induced subgraph of the UIR, dubbed \textit{atom dependency tree} (see \dinggreen{184} and \dingblue{184} in Figure~\ref{fig:rust_and_graphs}). This structure captures the \textit{lexical scope}-based~\cite{Cooper22} dependencies between atoms, enriched with the extent of the code they affect.
To address the previously mentioned limitations, we propose a method that combines both structures to identify the most relevant configurations.
This method leverages centrality measures as structural metrics~\cite{Fenton91, McCabe76, Newman10}, and includes graph transformations into propositional and conjunctive normal form (CNF)~\cite{Classen13} formulas, along with techniques for refining the ranking of features.
\ourtool{RustyEx} relies on a SAT solver to determine configurations that satisfy the CNF formula based on the refined feature ranking. This ensures that not only is the number of configurations reduced, but also that the most relevant configurations are prioritized.
We validate our approach on higher-ranked open-source Rust projects by running \ourtool{RustyEx} with a fixed number of configurations to generate.
To provide a comprehensive evaluation, we report detailed metrics on the proposed structures for each project.
Soundness is ensured by construction, guaranteeing that all generated configurations are valid.

\smallskip\noindent\textbf{Contributions.}\quad
The main contributions of this paper are:
\begin{compactitem}
   \item the first general method for prioritizing configurations in highly configurable software systems via a compiler-based refined ranking of features,
   \item a formalization of the approach, from the extraction of the UIR to the algorithms for building the complementary data structures and prioritizing configurations,
   \item a detailed implementation of our approach in \ourtool{RustyEx}, a fully automated tool for Rust software,
   \item an extensive evaluation of \ourtool{RustyEx} on higher-ranked open-source Rust projects, demonstrating its effectiveness in identifying and prioritizing the most relevant configurations, and
   \item a formal proof of the soundness of our approach, ensuring that all generated configurations are valid.
\end{compactitem}

\smallskip\noindent\textbf{Manuscript Structure.}\quad
The remainder of the paper is organized as follows. Section~\ref{sect:bg} introduces the necessary background.
Section~\ref{sect:design} details the design of \ourtool{RustyEx}, and Section~\ref{sect:evaluation} presents our evaluation. Section~\ref{sect:related-work} discusses related work, and Section~\ref{sect:conclusion} concludes the paper.

\section{Background}\label{sect:bg}
We provide background on Rust and its ownership system and on centrality measures.

\subsection{The Rust Programming Language.}\label{subsect:bg:rust}
Rust is a system programming language that focuses on safety, speed, and concurrency. It ensures memory safety without garbage collection, meaning pure Rust programs are free from null pointer dereferences and data races.
Rust's ownership system—integrated into its type system—is inspired by
\textit{linear logic}~\cite{Girard87, Girard95} and \textit{linear types}~\cite{Wadler90, Odersky92}, enforcing that each piece of memory (a value) has a single \textit{owner} (the variable binding) at any time~\cite{Clarke98, Boyapati02}.
When the owner goes out of scope, the memory is automatically deallocated, enabling user-defined destructors and supporting the \textit{resource acquisition is initialization} (RAII) pattern~\cite{Stroustrup94}.
Ownership can be transferred (\textit{moved}) or temporarily shared (\textit{borrowed}) through references. Rust supports two types of borrows: multiple \textit{immutable} borrows or a single \textit{mutable} borrow, but never both at the same time. These constraints, enforced by the compiler through \textit{borrow checking}, guarantee memory safety and prevent dangling pointers by ensuring that reference lifetimes never outlive their owners.
To support low-level operations, Rust provides \inlinerust{unsafe} blocks, where the compiler's safety guarantees are suspended and the burden of avoiding undefined behavior falls on the programmer. Outside of these blocks, Rust enforces strict safety, making undefined behavior impossible in safe code.
Rust supports software variability through its attribute system, particularly the \inlinerust{cfg} attribute, which allows conditional compilation based on specified configuration predicates. A feature in Rust is defined by all \textit{terms} (see footnote 1) annotated with the same \textit{config name} in a \inlinerust{cfg} \textit{outer attribute}.
The inclusion of a feature in a product depends on the evaluation of a \textit{configuration predicate} at compile time, e.g., \inlinerust{#[cfg(any(unix, windows))]}. 
\texttt{Cargo}---Rust's package manager---lets developers declare features using \textit{config names} in the \texttt{Cargo.toml} file, combining them with logical operators into \textit{configuration predicates}.
Feature dependencies, i.e., \textit{cross-tree constraints} of \textit{feature models}~\cite{Kang90,Seidl16,Karatas10}, can also be specified in the same file.

\subsection{Centrality Measures}\label{subsect:bg:centrality}
Since the early days of social network analysis, \textit{centrality measures} have been used in sociology and psychology to identify the key nodes in a network~\cite{Seeley49, Bavelas50, Katz53}.
They have since become central to graph theory and network analysis~\cite{Borgatti05, Das18, Landherr10}.
A \textit{network} is typically modeled as a graph with \(n\) nodes, indexed by \(i \in \{1, 2, \ldots, n\}\), and represented as an adjacency matrix \(A \in \mathbb{R}^{n \times n}\), where \(A_{ij}\neq0\) indicates an edge between nodes \(i\) and \(j\), and \(A_{ij}=0\) otherwise.
As Bloch \textit{et al.}~\cite{Bloch23} observe, not all centrality measures consider edge directions and weights, though these can be incorporated.
Formally, a centrality measure is a function \(\phi: \mathbb{R}^{n \times n} \rightarrow \mathbb{R}^n\), where \(\phi_i(A)\) quantifies the relevance of node \(i\) in the network \(A\).
As a cardinal invariant, it enables ordinal ranking of nodes based on their scores (\(\phi_i(A)\)).
Centrality measures are broadly categorized as: \textit{geometry-based} and \textit{spectral-based}~\cite{Boldi14}.
In the remainder of this section, we introduce some of the most commonly used centrality measures.

\smallskip\noindent\textbf{Geometric Centrality Measures.}\quad
The \textit{geometric} centrality measures assign relevance by using a function of distances---i.e., the number of nodes at each distance from a given node.
\textit{Degree centrality} counts the number of edges connected to a node: \(\textrm{d}^{-}(i)\) for in-degree and \(\textrm{d}^{+}(i)\) for out-degree.
\textit{Closeness centrality}, introduced by Bavelas~\cite{Bavelas50}, defines a node's relevance as inversely proportional to the sum of its distances node to all other nodes.
Since the original formulation fails on disconnected graphs (due to infinite distances), a widely adopted variant is: \[\textrm{Clo}_i(A) = \frac{1}{\underset{d(i, j)<\infty}{\sum_{j=1}^{n}}d(i, j)}\] where \(d(i, j)\) denotes the shortest path between nodes \(i\) and \(j\).
\textit{Harmonic centrality}~\cite{Marchiori00} is defined as the sum of the reciprocals of the shortest path lengths between a node and all others: \[\textrm{Har}_i(A) = \sum_{j=1}^{n}\frac{1}{d(i, j)}\]
Unlike closeness centrality, it is well-defined for disconnected graphs; for isolated nodes, the sum evaluates to \(0\) since \(d(i, j) = \infty\) for all \(j \neq i\).
\textit{Betweenness centrality}~\cite{Freeman77} measures how often a node appears on the shortest paths between pairs of nodes: \[\textrm{Bet}_i(A) = \sum_{s\neq i\neq t}\frac{\sigma_{st}(i)}{\sigma_{st}}\] where \(\sigma_{st}\) is the number of shortest paths from \(s\) to \(t\), and \(\sigma_{st}(i)\) is the number of those paths passing through node \(i\).
Newman~\cite{Newman01} extended \textit{closeness centrality} to weighted networks using Dijkstra's algorithm to compute for shortest paths.
Opsahl \textit{et al.}~\cite{Opsahl10} further generalized shortest-path measures by combining edge weights and counts, making them applicable to both \textit{closeness} and \textit{betweenness} centrality.

\smallskip\noindent\textbf{Spectral Centrality Measures.}\quad
The \textit{spectral} centrality measures evaluate node relevance using the dominant left eigenvector of the graph's adjacency matrix.
\textit{Eigenvector centrality}~\cite{Bonacich72} assigns higher scores to nodes connected to other high-scoring nodes. It is defined as the eigenvector \(\boldsymbol{v}\) associated with the largest eigenvalue \(\lambda\) of the adjacency matrix \(A\): \[\lambda\boldsymbol{v} = A\boldsymbol{v}\]
If \(A\) is a \textit{stochastic matrix}, the dominant eigenvalue is \(1\).
However, the \textit{eigenvector centrality} performs poorly on disconnected graphs~\cite{Berman79}.
\textit{Katz centrality}~\cite{Katz53} addresses this by considering the number of paths of varying lengths that connect a node to all other nodes. It is defined as: \[\boldsymbol{k} = \boldsymbol{1}{(I - \beta A)}^{-1}\] where \(\boldsymbol{1}\) is a vector of ones, \(I\) is the identity matrix, and \(\beta\) is a damping factor satisfying \(\beta<1/\lambda\), with \(\lambda\) the dominant eigenvalue of \(A\).

\section{A Deep Dive into \ourtool{RustyEx}}\label{sect:design}
As introduced in Section~\ref{sect:intro}, \ourtool{RustyEx} is a fully automated tool that instruments Rust compiler to: \begin{inparaenum}
   \item extract feature dependencies in Rust software,
   \item assign feature weights based on their impact on the code,
   \item apply centrality measures to rank features, and
   \item prioritize configurations.
\end{inparaenum}

\subsection{Process Overview}\label{subsect:design:architecture}
\begin{figure}[t]
   \centering
   \caption{The phases of \ourtool{RustyEx} (inspired from~\cite{Li21b})}%
   \includegraphics[width=\linewidth]{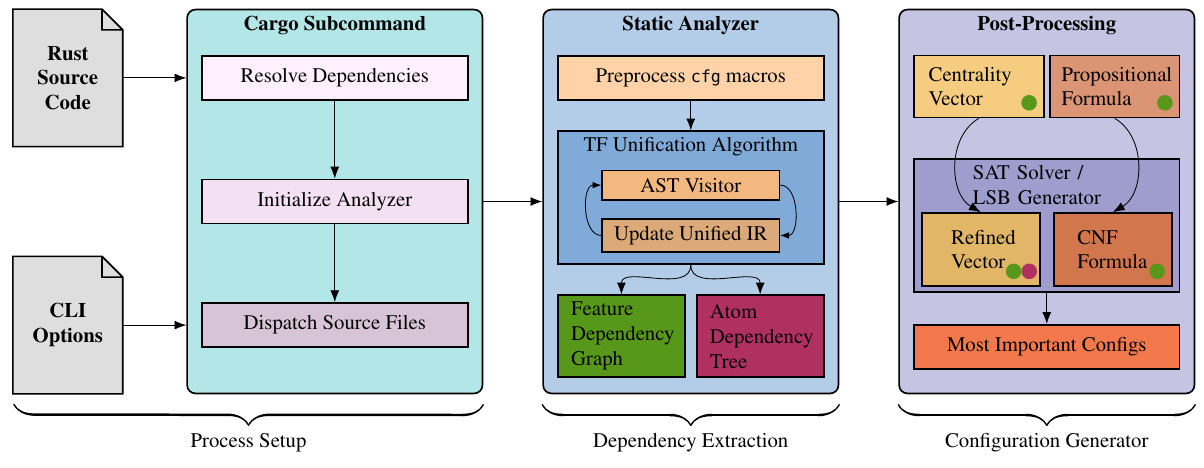}%
   \label{fig:architecture}
\end{figure}

\ourtool{RustyEx} performs its analysis in three main phases: \begin{inparaenum}
   \item \hlc[MyTealBlue!30]{process setup},
   \item \hlc[MyRoyalBlue!30]{dependency extraction}, and
   \item \hlc[MyViolet!30]{configuration generation},
\end{inparaenum}
as shown in Figure~\ref{fig:architecture}.

\smallskip\noindent\textbf{Process Setup.}\quad
Rust projects make extensive use of \inlinerust{cfg} attributes for conditional compilation, both for \textit{feature gating}---enabling or disabling features via \texttt{Cargo}---and for purposes like platform-specific code selection~\cite{Anderson16, Chen22, Li24}.
To ensure usability, \ourtool{RustyEx} integrates seamlessly with the Rust toolchain.
In this phase, it accepts a Rust project along with CLI options that customize the analysis, such as the number (\(N\)) of configurations to generate and the chose centrality measure.
The setup uses \texttt{Cargo} to resolve internal crate dependencies which are essential for the analysis.\footnote{%
   For performance, \ourtool{RustyEx} analyzes only the target crate's source files and compiles external dependencies with the standard Rust compiler. This reduces computational overhead but may slightly impact accuracy~\cite{Venkatakeerthy20}.%
}
It then initializes the analysis in memory based on the provided options and configures the dispatch of source files to use the instrumented \texttt{rustc}.

\smallskip\noindent\textbf{Dependency Extraction.}\quad
\ourtool{RustyEx} instruments the Rust compiler to extract the \textit{feature dependency graph} (see \dinggreen{183}, and \dingblue{183} in Figure~\ref{fig:rust_and_graphs}) and the \textit{atom dependency tree} (see \dinggreen{184}, and \dingblue{184} in Figure~\ref{fig:rust_and_graphs}) from the UIR derived from the AST\@. The details of these structures and their extraction are discussed in Section~\ref{sect:uir}.
This is integrated into the compiler by implementing the \inlinerust{rustc_driver_impl::Callback} trait, which provides four hooks: \inlinerust{config}, \inlinerust{after_crate_root_parsing}, \inlinerust{after_expansion}, and \inlinerust{after_analysis}.\footnote{See \url{https://doc.rust-lang.org/nightly/nightly-rustc/} for details.}
\texttt{rustc} performs the \textit{eager} macro expansion\footnote{
   Eager expansion expands macro arguments as early as possible, regardless of the macro invocation.
} after AST creation and before name resolution.
Although, \inlinerust{after_crate_root_parsing} would be ideal for analysis before macro expansion, it lacks access to sub-modules that have not been resolved yet. Thus, we hook into \inlinerust{after_expansion}, which is invoked after \textit{eager} macro expansion, name resolution, and AST validation.
To avoid premature evaluation of \inlinerust{cfg} attributes, we register a custom file loader in the \inlinerust{config} callback. This loader renames \inlinerust{cfg} macros to delay their evaluation.
In the \inlinerust{after_expansion} callback, we perform inter-procedural analysis on the AST to extract the UIR weighting its nodes with the \textit{weighted} fixed-point algorithm shown in Algorithm~\ref{alg:uir_weight} described in Section~\ref{sect:uir}.
These weights estimate each feature's impact on the code.

\smallskip\noindent\textbf{Configuration Generation.}\quad
Once the weighted feature dependency graph and the atom dependency tree are extracted, \ourtool{RustyEx} uses them to identify the most relevant configurations. The weighted feature dependency graph serves two key tasks: \begin{inparaenum}
   \item ranking features via \textit{geometry-based} and/or \textit{spectral-based} centrality measures (top-left box in the post-processing phase of Figure~\ref{fig:architecture}), and
   \item generating a corresponding propositional formula (top-right box in the post-processing phase of Figure~\ref{fig:architecture}).
\end{inparaenum}
These tasks are detailed in Section~\ref{sect:fdg}.
The propositional formula is then converted to CNF\@. Next, the atom dependency tree refines the feature ranking (middle boxes in the post-processing phase of Figure~\ref{fig:architecture}).
Finally, a SAT solver selects the top \(N\) configurations that satisfy the CNF formula, prioritizing those with the highest refined ranking---i.e., the most relevant configurations (bottom box in the post-processing phase of Figure~\ref{fig:architecture}).

\subsection{From AST to UIR}\label{sect:uir}

\begin{figure}[t]
   \centering
   \caption{\ourtool{RustyEx} process demonstrated on two scenarios}%
   \includegraphics[width=\linewidth]{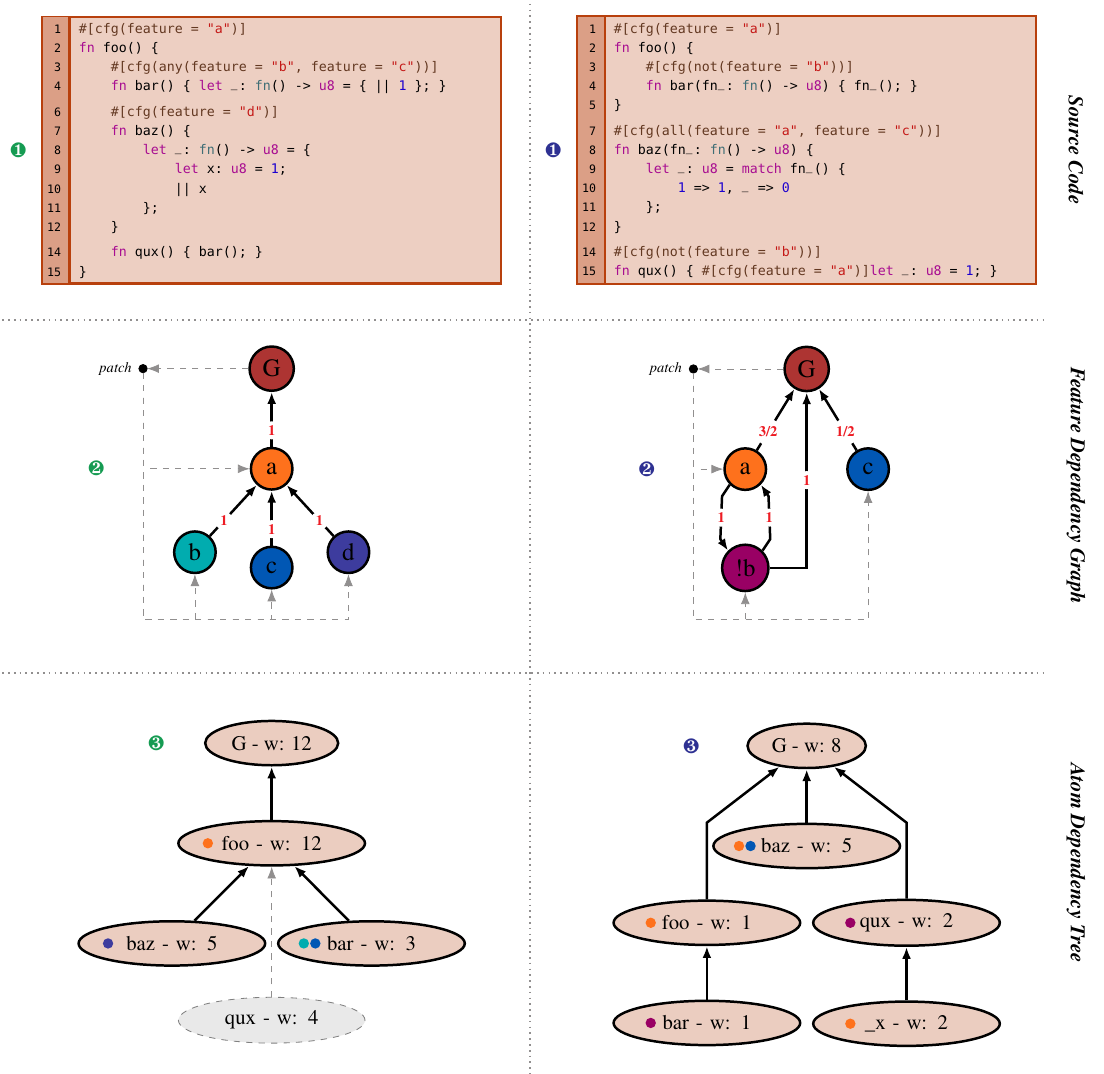}%
   \label{fig:rust_and_graphs}
\end{figure}

\noindent\textbf{Overview.}\quad
In the AST, the \inlinerust{cfg} attribute nodes and their associated \textit{terms} appear as separate child nodes under a common ancestor.
Building the UIR involves two key steps: \textit{unification} and \textit{weighting}.
The term UIR stems from the \textit{unification} step, which merges each \inlinerust{cfg} attribute with its associated \textit{term} into a single node called an \textit{atom}. UIR nodes are algebraic data types~\cite{Lehmann81, Pierce02, Turner85, Kennedy05, Bergstra95}, specifically \(\Sigma\)-types---i.e., nodes can be either \textit{atoms} or \textit{relevant} plain AST nodes.
Plain AST nodes are included in the UIR only if the \textit{weighting} step assigns them non-zero relevance---e.g., generic bounds on parameters are usually excluded.
For instance, in \dinggreen{182} and \dingblue{182} of Figure~\ref{fig:rust_and_graphs}, the function \inlinerust{foo} and its \inlinerust{#[cfg(feature = "a")]} attribute on line~1 are unified into a single \textit{atom} node, centered on the annotated item \inlinerust{foo}. Conversely, the body nodes of the function \inlinerust{bar} remain plain AST nodes in the UIR because they contribute to its weight, helping quantify how much of the code is influenced by the involved \inlinerust{cfg} features. 

\smallskip\noindent\textbf{Formalization.}\quad
We define the AST as a pair \((N_{ast}, E_{ast})\), where \(N_{ast}\) and \(E_{ast}\) are the sets of nodes and edges, respectively.
Let \(N_{rel}\subseteq N_{ast}\) denote the set of \textit{relevant} nodes.
A node is considered \textit{relevant} if it contributes to the semantics of a feature-dependent element. For instance, \inlinerust{let} statements are relevant, while \inlinerust{extern crate} or \inlinerust{use} declarations are not.

The set of \textit{atoms} is defined as: \[A = \{(p, t) \mid ann(t) = p \land t \in T \land p \in P\}\] where:\smallskip
\begin{compactenum}
   \item \(P\) is the set of all \inlinerust{cfg} predicates, each defined as a recursive \(\Sigma\)-type with the following structure:
   \begin{align*}
      p = \begin{cases}
             \texttt{single}(f)     & \text{for } f \in F        \\
             \texttt{not}(f)        & \text{for } f \in F        \\
             \texttt{any}(p_1, p_2) & \text{for } p_1, p_2 \in P \\
             \texttt{all}(p_1, p_2) & \text{for } p_1, p_2 \in P
          \end{cases}
   \end{align*}
   \item \(T \subseteq N_{ast}\) is the set of all \textit{terms}, and
   \item \(ann : T \rightarrow P\) is the surjective function that maps terms to \inlinerust{cfg} predicates.
\end{compactenum}
The UIR is an \textit{enhanced} induced subgraph of the AST, where nodes are enriched with \inlinerust{cfg} predicates and edge directions are reversed.
Formally, the UIR is defined as \(\mathcal{U} = (N, E, w_N)\), where:\smallskip
\begin{compactitem}
   \item \(N = A \ \cup \ N_{rel}\) with \(A\cap N_{rel}=\emptyset\), contains both \textit{atoms} and \textit{relevant} AST nodes, forming a \(\Sigma\)-type structure,
   \item \(E=\{(i, j)\mid(i, j)\in E'\}\), where \(E'\subseteq E_{ast}\) is the set of reversed edges derived from the AST, and
   \item \(w_N : N \rightarrow \mathbb{N}^{+}\) assigns weights to nodes based on the type of \textit{term}.
\end{compactitem}
Since UIR nodes may be compound \(\Sigma\)-types (e.g., atoms), \(E' \subseteq E_{ast}\) alone does not capture all necessary structure.
To preserve connectivity, if \((i, j) \in E_{ast}\) and there exists \(a \in A \subseteq N\) s.t.\ \(i = term(a)\) or \(j = term(a)\), we extend \(E'\) as: \[E'\leftarrow E'\cup\{\,((ann(i),\,i),\, j) \vee (i,\,(ann(j),\,j))\,\},\]
Here \(term:A\rightarrow T\) extracts the term from an \textit{atom}.
This ensures that when a node is unified into an \textit{atom}, its original AST edges are preserved in the UIR, as if the node remained \textit{relevant} on its own.

\smallskip\noindent\textbf{Unification Algorithm.}\quad
The \textit{unification algorithm} performs a \textit{depth-first visit} traversal of the AST to identify \inlinerust{cfg} attributes and their associated \textit{terms}, merging them into \textit{atoms}.
\ourtool{RustyEx} implements the visitor pattern~\cite{Gamma95} via the \inlinerust{rustc_ast::visit::Visitor} trait. To track parent-child relationships during traversal, it maintains a \textit{term stack}, where each visited term is pushed.
Since \inlinerust{cfg} attributes always appear as leftmost children of terms, encountering one triggers the \textit{unification} process. This involves:
\begin{compactenum}
   \item extracting the configuration predicate from the \inlinerust{cfg} attribute,
   \item parsing the predicate into a custom internal representation, dubbed \inlinerust{ComplexFeature}, and
   \item popping the corresponding term from stack to \textit{unify} it with the predicate into an \textit{atom} node.
\end{compactenum}
For example, the \inlinerust{cfg} attribute on line 3 of \dinggreen{182} of Figure~\ref{fig:rust_and_graphs} is parsed as
\begin{center}
   \showrust[.5\textwidth]{complex_feature.rs}
\end{center}
\noindent and unified with the term \inlinerust{bar} to form the corresponding \textit{atom} node.
Once a term is fully visited, it is popped from the stack and an UIR edge is added from the child to its parent.
Unlike the AST, UIR edges are reversed to reflect lexical scope-based relationships, capturing dependency flow---key for applying graph centrality measures in the feature dependency graph (Section~\ref{sect:fdg}).
The \inlinerust{bar} node in the atom dependency tree (\dinggreen{183} in Figure~\ref{fig:rust_and_graphs}) shows its unified features, highlighted by the \tikz\fill[MyTealBlue] (0,0) circle (0.1cm); and \tikz\fill[MyRoyalBlue] (0,0) circle (0.1cm); markers, emphasizing both unification and edge direction.

\begin{algorithm}[tbh!]
  \caption{Calculate The Weight of the UIR Nodes}%
  \label{alg:uir_weight}
  \begin{algorithmic}[1]\small
    \State $m_w \gets \emptyset$ 
    \State $Q \gets \emptyset$ 
    \State $r \gets root(\mathcal{U})$ 
    \State $\mathcal{U}' \gets \mathcal{U}^T$ \Comment{\textcolor{gray}{\small transpose}}
    \State \textsc{calc\_weight}($\mathcal{U}'$, $r$, $Q$, $m_w$)
    \State \textsc{resolve\_queue}($\mathcal{U}'$, $Q$, $m_w$)
    \State $\mathcal{U} \gets \mathcal{U}'^T$ \Comment{\textcolor{gray}{\small transpose back}}

    \vspace{1em}

    \Function{calc\_weight}{$\mathcal{U}'$, $n$, $Q$, $m_w$}
      \ForAll{$adj \in $ \textsc{adj}($\mathcal{U}', n$)} \Comment{\textcolor{gray}{\small adjacency nodes}}
        \State \textsc{calc\_weight}($\mathcal{U}'$, $adj$, $Q$, $m_w$)
      \EndFor

      \State $ch\_w \gets 0$ \Comment{\textcolor{gray}{\small children weight}}

      \ForAll{$adj \in $ \textsc{adj}($\mathcal{U}', n$)}
        \If{$adj.status =$ \textsc{Wait}} \Comment{\textcolor{gray}{\small check the status}}
          \State $n.status \gets$ \textsc{Wait}; \textbf{ret}
        \EndIf
        \State $ch\_w \gets ch\_w + adj.weight$
      \EndFor

      \State \textbf{match} $n.weight\_kind$ \textbf{with}
        \State \hspace{12pt} \textbf{case} \textsc{No}: $n.weight \gets 0$
        \State \hspace{12pt} \textbf{case} \textsc{Intrinsic}: $n.weight \gets 1 + ch\_w$
        \State \hspace{12pt} \textbf{case} \textsc{Children}: $n.weight \gets ch\_weight$
        \State \hspace{12pt} \textbf{case} \textsc{Reference}($called$):
          \State \hspace{24pt} \textbf{if} $m_w[called] = \emptyset$ \textbf{then}
          \Comment{\textcolor{gray}{\small no defs found}}
            \State \hspace{36pt} $n.status \gets$ \textsc{Wait}
            \State \hspace{36pt} $Q \gets Q \cup \{n\}$
            \State \hspace{36pt} \textbf{return}
          \State \hspace{24pt} \textbf{else}
          \State \hspace{36pt}$t.weight \gets$ \textsc{avg}($m_w[called]$)
          \State \hspace{24pt} \textbf{end if}
     \State \textbf{end match}
      \State $n.status \gets$ \textsc{Weighted}
      \State $m_w[n] \gets m_w[n] \cup {n.weight}$
    \EndFunction

    \vspace{1em}

    \Function{resolve\_queue}{$\mathcal{U}'$, $Q$, $m_w$}
      \State $S \gets \emptyset$ \Comment{\textcolor{gray}{\small set of seen nodes}}
      \While{$Q \neq \emptyset$}
        \State $c \gets Q\downarrow$
        \State $Q \gets Q - \{c\}$ \Comment{\textcolor{gray}{\small dequeue}}
        \If{$S \cap \{c, \textsc{len}(Q)\} \neq \emptyset$} \Comment{\textcolor{gray}{\small recovery mechanism}}
          \State $c.weight \gets$ DEF\_W \Comment{\textcolor{gray}{\small assign a default weight}}
          \State \textbf{continue}
        \EndIf

        \State $S \gets S \cup \{c, \textsc{len}(Q)\}$
        \State \textsc{calc\_weight}($\mathcal{U}'$, $c$, $Q$, $m_w$)
        \If{$c.status =$ \textsc{Wait}} 
          \State $Q \gets Q \cup \{c\}$
        \EndIf
      \EndWhile
    \EndFunction

  \end{algorithmic}
\end{algorithm}

\smallskip\noindent\textbf{Weighting Algorithm.}\quad
Each node in the UIR is assigned a positive integers, i.e., elements of \(\mathbb{N}^{+}\).
During the \textit{unification} phase, \ourtool{RustyEx} assigns to every UIR node a \textit{weight kind}, based solely on the kind of \textit{term} it represents. These \textit{weight kinds} determine how weights are computed in Algorithm~\ref{alg:uir_weight} (lines 16--25), and include:\smallskip
\begin{compactenum}
   \item \textit{no weight}: nodes irrelevant to the analysis that receive no weight (e.g., \inlinerust{extern crate} declarations),
   \item \textit{intrinsic weight}: nodes assigned a fixed weight of one (e.g., \inlinerust{let} statements),
   \item \textit{children weight}: nodes whose weight is the sum of their children weights (e.g., \inlinerust{foo} in \dinggreen{182} of Figure~\ref{fig:rust_and_graphs}), and
   \item \textit{reference weight}: nodes that refer to other nodes in the UIR and inherit their weight (e.g., the \inlinerust{bar()} call in \inlinerust{qux} in \dinggreen{182} of Figure~\ref{fig:rust_and_graphs}). 
\end{compactenum}
After \textit{unification}, \ourtool{RustyEx} runs a fixed-point algorithm (Algorithm~\ref{alg:uir_weight}) to compute the final weight function \(w_N\) over UIR nodes.
In Rust, multiple \textit{terms} can share the same identifier only if their \inlinerust{cfg} attributes are mutually exclusive. Let \(o_n\) denote the number of distinct \textit{atoms} a node \(n \in N\) appears in.
The algorithm initializes:\smallskip
\begin{compactitem}
   \item \(m_w : N \rightarrow {\mathbb{N}^{+}}^{o_n}\), a map storing all weights associated to the node identifier,
   \item an empty queue \(Q\),
   \item the root node \(r = root(\mathcal{U})\), and
   \item the transposed UIR \(\mathcal{U}' = \mathcal{U}^T\), since the original UIR is built bottom-up and transposing it reestablishes parent-child relationships.
\end{compactitem}
The \textit{weighting} algorithm starts at the root node \(r\), invoking the recursive \textsc{calc\_weight} function (line 17), which computes the weight of each node \(n \in N\) by traversing its children in \(\mathcal{U}'\) and applying the rules associated with its \textit{weight kind}.
Computed weights are stored both in the node and in the map \(m_w\) for reuse.
For each node \(n\in N\) of kind \textit{reference}, such as function calls, the algorithm checks whether the referenced node already has a computed weight in \(m_w\). If the weight is not yet available, the node is marked as \textit{wait} and added to the queue \(Q\).
The fixed-point \textsc{resolve\_queue} function (line 30) then processes \(Q\), computing weights for the queued nodes until the queue is empty or a cycle is detected---e.g., due to direct or mutual recursion~\cite{VonOheimb99, Magnusson07}.
A cycle is detected when \(Q\) remains unchanged between two complete iterations.
To prevents infinite loops, the algorithm assigns a default fallback weight to nodes involved in cycles and continues processing the remaining queue.\footnote{
   Another approach has been proposed by instrumenting LLVM-IR~\cite{Venkatakeerthy20}, attempting to solve a system of linear equations that, by definition, could not admit solutions.
}
Finally, the UIR is transposed back to its original direction (undoing the earlier transformation), and the finalized weights are stored in \(w_N\).


\subsection{From the UIR to the Feature Dependency Graph}\label{sect:fdg}

\smallskip\noindent\textbf{Overview.}\quad
Depicted in Figure~\ref{fig:rust_and_graphs} (\dinggreen{183}, \dingblue{183}), the feature dependency graph is a weighted directed graph that represents dependencies between software features. It is built by applying the \textit{feature-extraction} algorithm to the UIR\@. This algorithm defines rules for assigning edge weights based on the \textit{configuration predicates} of the \inlinerust{cfg} attributes in which the features appear.
At a high level, the graph reflects lexical-scope-based relationships among features. Specifically, if \(f_i\) is used within the lexical scope of an \textit{atom} annotated with \(f_j\), then \(f_i\) depends on \(f_j\) and a directed edge is created from \(f_i\) to \(f_j\).
For instance, in line 3 of Figure~\ref{fig:rust_and_graphs} \dinggreen{183}, \inlinerust{feature = "b"} and \inlinerust{feature = "c"} appear in an \inlinerust{any} \textit{configuration predicate}, within the lexical scope of an \textit{atom} unified with the \inlinerust{feature = "a"} (line 1) and enclosing the \inlinerust{foo} function (line 2). According to the \textit{feature-extraction} algorithm, two edges of weight \(1\) are created: from \inlinerust{feature = "b"} and from \inlinerust{feature = "c"} to \inlinerust{feature = "a"}. In contrast, at line 7 of \dingblue{183}, \inlinerust{feature = "c"} is used in an \inlinerust{all} \textit{configuration predicate} within the \textit{global} lexical scope (denoted as \hlc[MyRawSienna!60]{G}). Consequently, an edge of weight \(1/2\) is added from \inlinerust{feature = "c"} to \hlc[MyRawSienna!60]{G}. 
As in the UIR, edge directions in the feature dependency graph are reversed compared to the original AST\@.
The novel insight lies in preserving this natural edge direction: it allows us to interpret a feature's \textit{reputation} (or relevance) based on the \textit{state} (i.e., \textit{configuration predicate}) of the features that depend on it.
This enables the application of graph centrality measures on the feature dependency graph, where the edge weights are designed to support more accurate feature ranking.

\smallskip\noindent\textbf{Formalization.}\quad
Until now, we referred to the feature dependency graph as a graph. However, during its construction, it is initially a \textit{multigraph}, since a feature \(f_i\) may depend on another feature \(f_j\) in multiple parts of the code.
Formally, this multigraph is denoted as \(\mathcal{F} = (F, D, w_R)\), where:\smallskip
\begin{compactenum}
   \item \(F = \{(f_i, p_i) \mid f_i \in CN \land p_i \subseteq P\}\) is the set of nodes, where \(CN\) is the set of all \textit{feature names} and \(p_i\) is the set of all \inlinerust{cfg} predicates in which \(f_i\) is used;
   \item \(D = (U, m)\) is the \textit{multiset} of directed edges representing feature dependencies, where \(U = \{(i, j) \mid i, j \in F\}\) is the set of edges and \(m : U \rightarrow \mathbb{N}^{+}\) is the \textit{multiplicity} function that counts how many times a dependency occurs; and
   \item \(w_R : R \rightarrow \mathbb{R}\), where \[R = \{((i,j),k)\mid (i,j)\in U,\;1\leq k\leq m((i,j))\},\] is the weight function that assigns a weight to each individual edge instance, indexed by the multiplicity index \(k\), with weights determined by the \textit{configuration predicates} in which the dependency arises.\smallskip
\end{compactenum}
The \textit{feature-extraction} algorithm then squashes these multiple edges into single ones by aggregating their weights, effectively transforming the multigraph \(\mathcal{F}\) into a simple graph \(\mathcal{F}' = (F, D', w'_R)\), where: \[D' = \{(i,j)\mid(i,j)\in U\}\] is a set of unique edges obtained by removing the multiplicities from \(D\), and the aggregated weight function is defined as: \[w'_R((i, j)) = \sum_{k=1}^{m((i, j))} w_R((i, j), k),\:\forall (i, j)\in D'.\] Here, the sum aggregates the weights of all instances of the edge \((i, j)\), with \(m(i,j)\) denoting the total number of such instances. Each individual edge instance \((i,j,k)\) contributes to the total weight assigned to the edge \((i,j)\) in the final graph~\(\mathcal{F}'\).

\smallskip\noindent\textbf{Feature-Extraction Algorithm.}\quad
The \textit{feature-extraction} algorithm builds the feature dependency graph \(\mathcal{F}'\) from the UIR\@. It begins by iterating over the UIR atoms \(A \subseteq N\). For each atom \(a \in A\), it retrieves its parent node \(\hat{a} \in \{x \mid (\hat{a}, x) \in E\;\land\;x \in A\}\).
The existence of \(\hat{a}\) and uniqueness are guaranteed for all \(a \in A\) such that \(a\,\neq\,\) \hlc[MyRawSienna!60]{G}, since the UIR is an induced subgraph of the AST, and every node in a tree has a unique parent. This follows from the fact that the UIR is an \textit{enhanced} induced subgraph of the AST\@. Concretely, \(\hat{a}\) is the nearest ancestor of \(a\) that is an \textit{atom}.
The algorithm creates a directed edge from each feature \(f \varpropto pred(a) = p_a\) to each feature \(\hat{f} \varpropto pred(\hat{a}) = p_{\hat{a}}\), where \(\varpropto\) is read as ``\textit{involved in}'' and \(pred : A \rightarrow P\) returns the configuration predicate that annotates each \textit{atom}. The predicate sets of the \textit{feature nodes} \(f\) and \(\hat{f}\) are updated as:\smallskip\\
\((f, p)\leftarrow(f, p \land pred(a))\) and \((\hat{f}, p')\leftarrow(\hat{f}, p' \land pred(\hat{a}))\).\smallskip\\
The weights of the \textit{multi-edges} are derived from the nested \inlinerust{cfg} predicate of \(a\).
For each feature node \(\hat{f}\) and each \((f, w)\in\chi(p_a, 1)\), the algorithm creates a weighted edge from \(f\) to \(\hat{f}\), using the recursively defined function \(\chi\):
\begin{align*}
   \chi(p_a, w) =
   \begin{cases}
      [(f, w)],                                             & \hspace*{-1.3cm} \text{if } p_a \equiv \texttt{single}(f) \lor \texttt{not}(f) \\
      \chi(p_1, 1) \oplus \chi(p_2, 1),                     & \text{if } p_a \equiv \texttt{any}(p_1, p_2)                                   \\
      \chi(p_1, d_{p_1}^{w}) \oplus \chi(p_2, d_{p_2}^{w}), & \text{if } p_a \equiv \texttt{all}(p_1, p_2)
   \end{cases}
\end{align*}
where \(d_p^w = {(w \times |\{f \varpropto p\}|)}^{-1}\) and the \(\oplus\) infix operator concatenates two lists.
For example, in \dinggreen{182} of Figure~\ref{fig:rust_and_graphs}, the \inlinerust{feature = "b"} (line 3) appears in an \inlinerust{any} predicate nested under both \inlinerust{feature = "a"} (line 1) and \inlinerust{foo} (line 2), so an edge with weight \(1\) is added from \inlinerust{feature = "b"} to the \inlinerust{feature = "a"}. 
As discussed in Section~\ref{sect:fdg}, this weighting scheme prioritizes features in \inlinerust{any} predicates over those in \inlinerust{all} predicates, reflecting the fact that \inlinerust{any} predicates are less restrictive (only one feature needs to be enabled), whereas \inlinerust{all} predicates require all involved features to be enabled simultaneously.
Finally, the graph is simplified by summing the weights of all edges with the same source and target nodes.

\smallskip\noindent\textbf{Centrality Measures.}\quad
Once the feature dependency graph is built, \ourtool{RustyEx} ranks features by relevance using a centrality measure specified via command-line argument (see Figure~\ref{fig:architecture}, marked with a \tikz\fill[Chartreuse4!90] (0,0) circle (0.1cm); inside the \textit{Centrality Vector} box).
Since \(\mathcal{F}\) is a disconnected, weighted, directed graph, some centrality measures are either undefined or behave unpredictably~\cite{Berman79}. For example, \textit{eigenvector centrality} requires a connected graph~\cite{Berman79} and both \textit{closeness centrality} and \textit{betweenness centrality} need adaptations to handle weighted directed graphs~\cite{Newman01, Opsahl10}.\footnote{
   \ourtool{RustyEx} uses \texttt{rustworkx}~\cite{Treinish22} to perform centrality measures.
}
To ensure connectivity, we introduce a synthetic \textit{patch} node (see \dinggreen{183} and \dingblue{183} in Figure~\ref{fig:rust_and_graphs}).
This node has a single incoming edge from \hlc[MyRawSienna!60]{G} and outgoing edges to all other nodes, each with weight \(1\).
This transformation guarantees that \(\mathcal{F}\) is fully connected, allowing the following centrality measures to be applied without modification:
\begin{compactitem}
   \item \textit{Newman Closeness Centrality}~\cite{Newman01}
   \item \textit{Opshal Betweenness Centrality}~\cite{Opsahl10}
   \item \textit{Eigenvector Centrality}~\cite{Bonacich72}
   \item \textit{Katz Centrality}~\cite{Katz53}
\end{compactitem}
After computing centrality scores, a vector \(\vec{v}\)  is produced where each entry \(v_i\) represents the centrality of feature \(f_i\). The \textit{patch} nodes are excluded from the vector, as they are not \textit{extracted} features.
This approach aligns with the intuition behind our graph construction: a feature’s reputation is defined by the features that depend on it. Centrality measures serve as a principled way to rank features by structural relevance.

\smallskip\noindent\textbf{Propositional Formula and CNF.}\quad
From the feature dependency graph, \ourtool{RustyEx} builds a \textit{propositional formula} by iterating over each feature node \(f_i \in F\).
For each node, it generates a clause \(\varphi_i = \ell(pred(f_i))\), where \(\ell\) maps \inlinerust{cfg} predicates to \textit{logical} expressions.
It then adds an implication \(f_i \Rightarrow f_j\) for each edge \((f_i, f_j)\in D'\), capturing feature dependencies.
The overall formula \(\varphi\) is defined as:
\[\varphi \;=\; \bigwedge_{f_i\in F} \Bigl(\varphi_i\;\wedge\;\bigwedge_{f_j\in \text{succ}(f_i)}(f_i\Rightarrow f_j)\Bigr)\]
where \(\text{succ}(f_i)\) is the set of successors of \(f_i\) in \(\mathcal{F}'\).
Finally, \(\varphi\) is converted to CNF formula \(\varphi_{cnf}\)~\cite{Howson97}. Duplicate clauses may be removed before or after the CNF conversion.

\subsection{From the UIR to the Atom Dependency Tree}\label{subsect:rusty-ex:adt}
\smallskip\noindent\textbf{Overview.}\quad
Figure~\ref{fig:rust_and_graphs} (boxes \dinggreen{184} and \dingblue{184}) shows the atom dependency tree derived from the UIR\@. The name highlights its selective retention of UIR \(\Sigma\)-type nodes: only the \textit{atom} variant is preserved, while \textit{plain} AST nodes are removed. Specifically, only UIR nodes with \inlinerust{cfg} attributes are kept. For instance, in Figure~\ref{fig:rust_and_graphs}, box \dinggreen{184}, the gray \inlinerust{qux} node (representing the \inlinerust{qux} function from line 14 in Figure~\ref{fig:rust_and_graphs}, box \dinggreen{182}) is excluded because it lacks any \inlinerust{cfg} attribute.
The atom dependency tree captures lexical scope-based dependencies among UIR \textit{atoms}. Unlike the feature dependency graph, which flattens UIR structure, this tree preserves parent-child relationships between \textit{atoms} and, by extension, between \textit{features}. Each node is weighted  by the amount of code affected by that atom's \inlinerust{cfg} condition, reflecting the feature's impact.
To avoid losing information when \textit{plain} AST nodes are discarded, the \textit{atom dependency extraction} algorithm aggregates their weights into their nearest ancestor \textit{atom} node (see \inlinerust{foo} node in Figure~\ref{fig:rust_and_graphs}, box \dinggreen{184}). Finally, the centrality vector \(\vec{v}\) is refined using the structure of the atom dependency tree to get a better ranking of feature relevance.

\smallskip\noindent\textbf{Formalization.}\quad
Given the UIR definition \(\mathcal{U} = (N, E, w_N)\), we define the atom dependency tree as a directed graph \(\mathcal{A} = (A, E_A, w_A)\), where:\smallskip
\begin{compactitem}
   \item \(A = N \mathbin{\raisebox{0.4ex}{\(\scriptscriptstyle\smallsetminus\)}} N_{rel}\) is the set of \textit{atom} nodes from the UIR,
   \item \(E_A = E \mathbin{\raisebox{0.4ex}{\(\scriptscriptstyle\smallsetminus\)}} \{(i, j) \mid (i, j) \in E \land \left( i \in N_{rel} \lor j \in N_{rel} \right)\}\) is the set of edges only connecting \textit{atom} nodes, and
   \item \(w_A : A \rightarrow \mathbb{N}^{+}\) is the weight function that assigns to each atom a weight reflecting the amount of code affected by the \inlinerust{cfg} attributes in which its features appear.
\end{compactitem}
The atom dependency tree (\(\mathcal{A}\)) is an induced subgraph of \(\mathcal{U}\), containing all \textit{atoms} and their dependencies, but excludes \textit{plain} AST nodes.
This holds trivially since \(A \subseteq N\) and \(E_A \subseteq E\).

\smallskip\noindent\textbf{Atom Dependency Extraction Algorithm.}\quad
The \textit{atom dependency extraction} algorithm builds the atom dependency tree from the UIR\@. It iterates over the UIR atoms \(A \subseteq N\). For each atom \(a \in A\), it identifies its parent \(\hat{a}\in\{x\mid(\hat{a}, x)\in E\;\land\;x\in A\}\). As before, the existence of \(\hat{a}\) is trivial to prove.
The algorithm then distinguishes two cases: \begin{inparaenum}
   \item if \(pred(a) \neq \emptyset\), it creates a edge from \(a\) to \(\hat{a}\) and updates \(\hat{a}\)'s weight by adding \(a\)'s weight,
   \item if \(pred(a) = \emptyset\), it \textit{does not} create an edge from \(a\) to \(\hat{a}\) but still updates \(\hat{a}\)'s weight. For example, in Figure\ref{fig:rust_and_graphs}, box \dinggreen{184}, the \inlinerust{qux} node contributes to the weight of \inlinerust{foo} but does not create an edge.
\end{inparaenum}
The parent weight update is performed as:\[w_A(\hat{a}) \leftarrow w_A(\hat{a}) + w_N(a).\]

\smallskip\noindent\textbf{Refinement Algorithm.}\quad
The \textit{refinement algorithm} refines the centrality vector \(\vec{v}\) using the atom dependency tree \(\mathcal{A}\) (see Figure~\ref{fig:architecture}).
It iterates over each node \(a \in A\) in \(\mathcal{A}\), extracts every feature \(f \varpropto pred(a)\), and updates its centrality value in \(\vec{v}\) as:
\begin{align*}
   \vec{v}_f & \leftarrow \vec{v}_f + \begin{cases}
                                         \overline{w_A(a)},                                                 & \hspace*{-1.3cm} \text{if } f \varpropto \texttt{single}(f)\;\lor\;f \varpropto \texttt{not}(f) \\
                                         \overline{w_A(a)},                                                 & \hspace*{.85cm} \text{if } f \varpropto \texttt{any}(p_1, p_2)                                  \\[2.5pt]
                                         \displaystyle\frac{\overline{w_A(a)}}{|\{f \varpropto pred(a)\}|}, & \hspace*{.85cm} \text{if } f \varpropto \texttt{all}(p_1, p_2)                                  \\
                                      \end{cases}
\end{align*}
Here \(f \varpropto\,p\), with \(p\in P\), means that \(f\) appears in the \(p\)-variant of \(a\)'s \inlinerust{cfg} predicate, and \(\overline{w_A(a)} \in [0, 1]\) is the normalized weight of the node \(a\). The normalization prevents the inclusion of the existing centrality values in \(\vec{v}\).
Although \(\mathcal{F}\) captures dependencies between features, it may fall short of expressing their actual relevance.
The atom dependency tree \(\mathcal{A}\) refines centrality values by incorporating the \textit{extent} of code affected by each feature, yielding \(\vec{v}'\) a new permutation of the vector~\(\vec{v}\).



\subsection{Configuration Generation}\label{sect:config_generation}
\textit{Configuration generation} is the final step of \ourtool{RustyEx}.
It produces the most relevant configurations based on the centrality vector \(\vec{v}'\) and the CNF formula \(\varphi_{cnf}\) (see Figure~\ref{fig:architecture}).
While more advanced lexical-scope-based strategies could be considered, \ourtool{RustyEx} uses SAT solvers---both incremental (e.g., \texttt{MiniSat}~\cite{Een03, Sorensson05} and \texttt{CaDiCaL}~\cite{Biere20}) and non-incremental (e.g., \texttt{Kissat}~\cite{Biere20}).
Given a fixed number \(K\) of configurations to generate, the algorithm iterates over the values in \(\vec{v}'\). For each feature \(f_i\), it updates the formula as:\[\varphi_{cnf} \leftarrow \varphi_{cnf} \land f_i\]
Then, it uses a SAT solver to generate all the configurations satisfying \(\varphi_{cnf}\).
A SAT solver is then invoked to produce all satisfying configurations for the updated formula. If a satisfying configuration is found, it is added to the result set. This process repeats until \(K\) configurations are obtained, negating each newly found configuration to favor the discovery of new ones.



\subsection{Applications and Clarifications}\label{sect:applications}


The generation of configurations in \ourtool{RustyEx} is performed incrementally and in a \textit{lazy} fashion. The tool ranks all features according to their refined centrality in $\vec{v}'$ and then generates only the top $K$ configurations most likely to cover critical feature interactions. In practice, this means prioritizing configurations that include the highest-ranked features in $\vec{v}'$, starting with $\vec{v}'_0$, the most central feature.

As discussed earlier, this prioritization strategy supports efficient exploration of the configuration space by focusing on variants that are most representative and impactful in terms of feature relevance. By analyzing these configurations first, developers can uncover important interactions and behaviors in the most relevant scenarios, improving overall system understanding and validation. Moreover, this targeted approach significantly reduces the computational cost and resource consumption associated with exhaustive analysis of all configurations, making it a practical solution for large-scale, highly configurable systems. For instance, these prioritized configurations could also serve as candidates for targeted performance profiling or static analysis, ensuring that critical feature combinations are examined before less relevant ones.

This configuration-aware pipeline highlights the potential of centrality-guided heuristics in feature-oriented analysis of configurable systems and paves the way for future work in optimizing configuration sampling based on structural and semantic program properties. It suggests a broader paradigm where program structure and semantics guide systematic exploration of the configuration space, potentially benefiting tasks such as performance evaluation or security analysis.


\afterpage{%
   \clearpage
   \begin{landscape}
      \begin{table*}[tbh!]
  \caption{Results of the experiments conducted on 40 Rust projects. All columns are aggregated per project, summing the values of all crates within the workspace, except for \textit{Peak Memory Usage} and \textit{UIR Height}, which are the maximum values.}%
  \label{table:results}
  \centering
  \resizebox{1.25\textwidth}{!}{%
  \begin{tabular}{>{\ttfamily}lrrrrrrrrrrrrrrrrrrrrrrrrrrrrrrrr}
    \toprule
    \multicolumn{1}{>{\bfseries\scriptsize\centering}p{1cm}}{Project Name} &
    \multicolumn{1}{>{\bfseries\scriptsize\centering}p{1cm}}{GitHub Stars} &
    \multicolumn{1}{>{\bfseries\scriptsize\centering}p{1cm}}{Crates.io\\Downloads} &
    \multicolumn{1}{>{\bfseries\scriptsize\centering}p{1cm}}{Lines of Code} &
    \multicolumn{1}{>{\bfseries\scriptsize\centering}p{1cm}}{Workspace\\Members} &
    \multicolumn{1}{>{\bfseries\scriptsize\centering}p{1cm}}{Failed Members} &
    \multicolumn{1}{>{\bfseries\scriptsize\centering}p{1cm}}{Dependency} &
    \multicolumn{1}{>{\bfseries\scriptsize\centering}p{1cm}}{Defined Features} &
    \multicolumn{1}{>{\bfseries\scriptsize\centering}p{1cm}}{UIR Nodes} &
    \multicolumn{1}{>{\bfseries\scriptsize\centering}p{1cm}}{UIR Edges} &
    \multicolumn{1}{>{\bfseries\scriptsize\centering}p{1cm}}{UIR Height} &
    \multicolumn{1}{>{\bfseries\scriptsize\centering}p{1cm}}{Feat.~D.G. Nodes} &
    \multicolumn{1}{>{\bfseries\scriptsize\centering}p{1cm}}{Feat.~D.G. Edges} &
    \multicolumn{1}{>{\bfseries\scriptsize\centering}p{1cm}}{Squashed D.G.~Edges} &
    \multicolumn{1}{>{\bfseries\scriptsize\centering}p{1cm}}{Atom D.T.~Nodes} &
    \multicolumn{1}{>{\bfseries\scriptsize\centering}p{1cm}}{Atom D.T.~Edges} &
    \multicolumn{1}{>{\bfseries\scriptsize\centering}p{1cm}}{Execution Time} &
    \multicolumn{1}{>{\bfseries\scriptsize\centering}p{1cm}}{Memory Usage} \\\midrule

    \href{https://github.com/rustdesk/rustdesk}{{rustdesk}}                        & 81965            & 2826               & 108908           & 8             & 1            & 67             & 8             & 1353             & 1346             & 13           & 24            & 41             & 27            & 29            & 22            & 1\,s              & 53\,MB            \\
    \href{https://github.com/GitoxideLabs/gitoxide}{{gitoxide}}                    & 9490             & 76429              & 223296           & 79            & 2            & 554            & 139           & 180166           & 180089           & 53           & 321           & 910            & 473           & 587           & 510           & 817\,s            & 999\,MB           \\
    \href{https://github.com/denoland/deno}{{deno}}                                & 101658           & 419085             & 284483           & 36            & 3            & 561            & 8             & 123883           & 123850           & 31           & 124           & 1220           & 153           & 570           & 537           & 320\,s            & 945\,MB           \\
    \href{https://github.com/tauri-apps/tauri}{{tauri}}                            & 89324            & 3840655            & 82089            & 26            & 2            & 255            & 65            & 26175            & 26151            & 30           & 78            & 193            & 89            & 144           & 120           & 108\,s            & 902\,MB           \\
    \href{https://github.com/FuelLabs/sway}{{sway}}                                & 62453            & 1092               & 210721           & 28            & 4            & 369            & 7             & 119724           & 119700           & 33           & 75            & 118            & 78            & 91            & 67            & 1156\,s           & 1910\,MB          \\
    \href{https://github.com/FuelLabs/fuel-core}{{fuel-core}}                      & 57814            & 265047             & 148682           & 36            & 4            & 370            & 69            & 85528            & 85496            & 31           & 132           & 437            & 188           & 351           & 319           & 153\,s            & 488\,MB           \\
    \href{https://github.com/alacritty/alacritty}{{alacritty}}                     & 57640            & 192714             & 32812            & 5             & 1            & 39             & 6             & 27244            & 27240            & 24           & 17            & 47             & 25            & 36            & 32            & 117\,s            & 513\,MB           \\
    \href{https://github.com/zed-industries/zed}{{zed}}                            & 54095            & 54231              & 591481           & 178           & 2            & 2458           & 91            & 23791            & 23615            & 28           & 381           & 268            & 237           & 231           & 55            & 91\,s             & 497\,MB           \\
    \href{https://github.com/sharkdp/bat}{{bat}}                                   & 51048            & 1496904            & 14971            & 1             & 0            & 30             & 9             & 675              & 674              & 19           & 5             & 14             & 7             & 8             & 7             & 26\,s             & 339\,MB           \\
    \href{https://github.com/BurntSushi/ripgrep}{{ripgrep}}                        & 50241            & 938719             & 50285            & 10            & 2            & 40             & 8             & 60181            & 60173            & 27           & 43            & 158            & 69            & 124           & 116           & 320\,s            & 887\,MB           \\
    \href{https://github.com/meilisearch/meilisearch}{{meilisearch}}               & 49152            & 1288               & 166851           & 19            & 2            & 196            & 30            & 44000            & 43983            & 30           & 46            & 69             & 41            & 55            & 38            & 404\,s            & 2477\,MB          \\
    \href{https://github.com/FuelLabs/fuels-rs}{{fuels-rs}}                        & 43907            & 5611               & 42484            & 24            & 2            & 90             & 22            & 43578            & 43556            & 28           & 70            & 165            & 78            & 138           & 116           & 161\,s            & 691\,MB           \\
    \href{https://github.com/typst/typst}{{typst}}                                 & 37358            & 57580              & 112435           & 20            & 4            & 185            & 15            & 48913            & 48897            & 27           & 49            & 79             & 50            & 57            & 41            & 81\,s             & 465\,MB           \\
    \href{https://github.com/helix-editor/helix}{{helix}}                          & 35745            & 103645             & 93042            & 14            & 1            & 143            & 12            & 55807            & 55794            & 28           & 59            & 175            & 89            & 133           & 120           & 595\,s            & 980\,MB           \\
    \href{https://github.com/charliermarsh/ruff}{{ruff}}                           & 35669            & 4764               & 370658           & 38            & 3            & 402            & 23            & 124392           & 124357           & 51           & 130           & 237            & 161           & 176           & 141           & 462\,s            & 504\,MB           \\
    \href{https://github.com/lapce/lapce}{{lapce}}                                 & 34961            & 35019              & 67802            & 5             & 2            & 73             & 3             & 20525            & 20522            & 41           & 16            & 44             & 25            & 32            & 29            & 195\,s            & 948\,MB           \\
    \href{https://github.com/nushell/nushell}{{nushell}}                           & 33821            & 975                & 291737           & 36            & 6            & 213            & 26            & 126096           & 126066           & 28           & 108           & 303            & 138           & 232           & 202           & 662\,s            & 956\,MB           \\
    \href{https://github.com/pola-rs/polars}{{polars}}                             & 31805            & 2400845            & 376212           & 26            & 3            & 362            & 789           & 58368            & 58345            & 32           & 103           & 448            & 161           & 354           & 331           & 277\,s            & 529\,MB           \\
    \href{https://github.com/swc-project/swc}{{swc}}                               & 31649            & 1767375            & 648247           & 116           & 15           & 957            & 230           & 412313           & 412212           & 258          & 326           & 1122           & 369           & 722           & 621           & 2016\,s           & 3425\,MB          \\
    \href{https://github.com/influxdata/influxdb}{{influxdb}}                      & 29458            & 283996             & 54304            & 19            & 2            & 352            & 15            & 62082            & 62065            & 27           & 58            & 116            & 67            & 92            & 75            & 164\,s            & 504\,MB           \\
    \href{https://github.com/TabbyML/tabby}{{tabby}}                               & 29742            & 5637               & 41457            & 19            & 1            & 270            & 15            & 52972            & 52954            & 21           & 55            & 134            & 56            & 109           & 91            & 442\,s            & 1827\,MB          \\
    \href{https://github.com/servo/servo}{{servo}}                                 & 29222            & 8792               & 367323           & 8             & 1            & 45             & 19            & 2670             & 2663             & 10           & 19            & 23             & 17            & 17            & 10            & 1\,s              & 64\,MB            \\
    \href{https://github.com/wasmerio/wasmer}{{wasmer}}                            & 19335            & 4820432            & 263877           & 35            & 8            & 350            & 153           & 72860            & 72833            & 26           & 121           & 318            & 164           & 235           & 208           & 246\,s            & 435\,MB           \\
    \href{https://github.com/diem/diem}{{diem}}                                    & 16702            & 16759              & 428331           & 186           & 10           & 2239           & 130           & 427311           & 427135           & 65           & 477           & 1207           & 475           & 757           & 581           & 1359\,s           & 968\,MB           \\
    \href{https://github.com/EmbarkStudios/texture-synthesis}{{texture-synthesis}} & 1768             & 61696              & 4735             & 3             & 0            & 7              & 2             & 13211            & 13208            & 23           & 15            & 30             & 22            & 21            & 18            & 11\,s             & 295\,MB           \\
    \href{https://github.com/EmbarkStudios/kajiya}{{kajiya}}                       & 5006             & 1060               & 26638            & 14            & 3            & 97             & 3             & 28922            & 28911            & 29           & 25            & 22             & 17            & 19            & 8             & 68\,s             & 1395\,MB          \\
    \href{https://github.com/EmbarkStudios/rust-gpu}{{rust-gpu}}                   & 7412             & 2135               & 44422            & 19            & 1            & 66             & 24            & 7581             & 7563             & 34           & 46            & 41             & 38            & 31            & 13            & 12\,s             & 360\,MB           \\
    \href{https://github.com/paritytech/substrate}{{substrate}}                    & 8381             & 1938               & 595987           & 270           & 8            & 2986           & 554           & 516169           & 515907           & 43           & 863           & 2172           & 1002          & 1681          & 1419          & 1135\,s           & 1801\,MB          \\
    \href{https://github.com/quickwit-oss/tantivy}{{tantivy}}                      & 12662            & 5253776            & 118896           & 9             & 1            & 67             & 11            & 34915            & 34907            & 51           & 32            & 84             & 41            & 66            & 58            & 104\,s            & 478\,MB           \\
    \href{https://github.com/hyperium/tonic}{{tonic}}                              & 10477            & 93238823           & 41456            & 29            & 0            & 166            & 42            & 37584            & 37555            & 25           & 88            & 330            & 104           & 254           & 225           & 270\,s            & 545\,MB           \\
    \href{https://github.com/n0-computer/sendme}{{sendme}}                         & 357              & 15473              & 1031             & 1             & 0            & 20             & 0             & 1577             & 1576             & 18           & 2             & 1              & 1             & 1             & 0             & 52\,s             & 525\,MB           \\
    \href{https://github.com/moghtech/komodo}{{komodo}}                            & 2766             & 822                & 63829            & 12            & 3            & 57             & 2             & 12443            & 12434            & 33           & 20            & 33             & 13            & 31            & 22            & 80\,s             & 685\,MB           \\
    \href{https://github.com/cloudflare/quiche}{{quiche}}                          & 9841             & 541561             & 84162            & 9             & 2            & 28             & 0             & 23455            & 23448            & 67           & 25            & 62             & 30            & 50            & 43            & 28\,s             & 352\,MB           \\
    \href{https://github.com/rolldown/rolldown}{{rolldown}}                        & 10119            & 941                & 49343            & 35            & 1            & 256            & 12            & 35014            & 34980            & 32           & 89            & 149            & 76            & 126           & 92            & 64\,s             & 354\,MB           \\
    \href{https://github.com/n0-computer/iroh}{{iroh}}                             & 3892             & 96831              & 39197            & 7             & 1            & 129            & 14            & 15758            & 15752            & 20           & 22            & 70             & 28            & 43            & 37            & 100\,s            & 381\,MB           \\
    \href{https://github.com/succinctlabs/sp1}{{sp1}}                              & 1194             & 1238               & 104219           & 25            & 1            & 363            & 44            & 70776            & 70752            & 53           & 77            & 183            & 87            & 147           & 123           & 248\,s            & 549\,MB           \\
    \href{https://github.com/unionlabs/union}{{union}}                             & 22086            & 11795              & 387611           & 148           & 2            & 1562           & 248           & 134318           & 134172           & 33           & 518           & 843            & 605           & 605           & 459           & 212\,s            & 563\,MB           \\
    \href{https://github.com/juspay/hyperswitch}{{hyperswitch}}                    & 13381            & 13567              & 575328           & 33            & 6            & 450            & 116           & 107187           & 107160           & 29           & 123           & 754            & 183           & 578           & 551           & 599\,s            & 3800\,MB          \\
    \href{https://github.com/emilk/egui}{{egui}}                                   & 23706            & 4248075            & 101376           & 39            & 4            & 138            & 44            & 39574            & 39539            & 44           & 111           & 235            & 126           & 181           & 146           & 88\,s             & 489\,MB           \\
    \href{https://github.com/Nukesor/pueue}{{pueue}}                               & 5297             & 72209              & 18340            & 3             & 2            & 24             & 0             & 15463            & 15462            & 23           & 5             & 18             & 8             & 15            & 14            & 64\,s             & 506\,MB           \\ \midrule

    \multicolumn{1}{>{\bfseries}p{1cm}}{Total}                                                                & \textbf{1212599} & \textbf{120362360} & \textbf{7329058} & \textbf{1628} & \textbf{116} & \textbf{17036} & \textbf{3008} & \textbf{3294554} & \textbf{3293042} & \textbf{258} & \textbf{4898} & \textbf{12873} & \textbf{5618} & \textbf{9129} & \textbf{7617} & \textbf{13328\,s} & \textbf{3800\,MB} \\ \midrule
    \multicolumn{1}{>{\bfseries}p{1cm}}{Average}                                                              & \textbf{30314}   & \textbf{3009059}   & \textbf{183226}  & \textbf{40}   & \textbf{2}   & \textbf{425}   & \textbf{75}   & \textbf{82363}   & \textbf{82326}   & \textbf{37}  & \textbf{122}  & \textbf{321}   & \textbf{140}  & \textbf{228}  & \textbf{190}  & \textbf{333\,s}   & \textbf{885\,MB}  \\ \bottomrule

  \end{tabular}
  }
\end{table*}

   \end{landscape}
   \clearpage
}

\section{Evaluation}\label{sect:evaluation}
The evaluation of our approach is twofold: \begin{inparaenum}
   \item we prove the soundness of \ourtool{RustyEx} by showing that the generated configurations are valid, and
   \item we evaluate the performance of \ourtool{RustyEx} on a set of \(40\) real-world, high-ranking open-source Rust projects.
\end{inparaenum}
\subsection{Soundness}\label{sect:soundness}
\smallskip\noindent\textbf{Premises.}\quad
To prove correctness, we show that: \begin{inparaenum}
   \item the CNF encoding faithfully represents all configuration constraints,
   \item the SAT solver returns only satisfying assignments, and
   \item no invalid configuration is ever generated.
\end{inparaenum}

\smallskip\noindent\textbf{Construction of the CNF Formula.}\quad
\textit{Clause formation}---for each feature node \(f_i \in F\), a clause \(\varphi_i = \ell(pred(f_i))\) is created to enforce the logical interpretation of the feature's configuration predicate. That is, if a feature is enables in some configuration, then its predicate must be satisfied.
\textit{Implication formation}---for every outgoing edge to \(f_j\) from \(f_i\) (i.e., for each \(f_j\) such that \((f_i, f_j)\in D'\)), an implication \(f_i \Rightarrow f_j\) is added to capture the dependency relations. This ensures that if feature \(f_i\) is active, then all its dependent \(f_j\) are also active.
\textit{Extension}---the formula \(\varphi\) is extended to include mandatory features specified in \texttt{Cargo.toml}, guaranteeing their presence in all generated configurations.
\textit{Equivalence Preservation}---since the CNF conversion uses standard techniques that preserve logical equivalence, any assignment that satisfying the CNF formula \(\varphi_{cnf}\) also satisfies the original formula \(\varphi\).

\smallskip\noindent\textbf{Correctness of the SAT Solution.}\quad
Given a SAT solver sound and complete,\footnote{A SAT solver is \textit{sound} if it only returns assignments that satisfy the formula, and \textit{complete} if it finds a solution whenever one exists.} any assignment \(\alpha\) returned for \(\varphi_{cnf}\) satisfies every clause in \(\varphi_{cnf}\), and thus also in \(\varphi\). This implies:\smallskip
\begin{compactenum}
   \item for each feature \(f_i\) with clause \(\varphi_i\), its activation under \(\alpha\) satisfies its configuration predicate: \(\alpha \models \varphi_i \quad \forall f_i \in F\)---that is, \(\alpha\) models the predicate;
   \item for each dependency \(f_i \Rightarrow f_j\), the implication holds under the assignment \(\alpha\) if \(\alpha(f_i) = \texttt{true} \Rightarrow \alpha(f_j) = \texttt{true}\);
   \item all mandatory features from \texttt{Cargo.toml} are active in \(\alpha\): \(\forall f_i \in F_{mandatory},\; \alpha(f_i) = \texttt{true}\).\smallskip
\end{compactenum}

\noindent By definition, a configuration is valid if and only if:\smallskip
\begin{compactitem}
   \item it satisfies all the configuration constraints including feature predicates and dependencies, and
   \item it includes all mandatory features.
\end{compactitem}

\noindent Therefore, since any assignment \(\alpha\) produced by the SAT solver satisfies all clauses in \(\varphi_{cnf}\), every configuration generated by \ourtool{RustyEx} is valid.

\smallskip\noindent\textbf{Reinforcement by Contradiction.}\quad
Assume, for contradiction, that the SAT solver returns a configuration \(\alpha'\) that is invalid---that is, it violates one or more constraints. Then there exists at least one clause in \(\varphi_{cnf}\) that is \textit{not} satisfied by \(\alpha'\). By definition, a \textit{satisfying assignment} satisfies all clauses in the formula, and the soundness of the SAT solver guarantees that \(\alpha'\) \textit{is} a satisfying assignment for \(\varphi_{cnf}\).
This contradiction implies that our assumption is false: \(\alpha'\) cannot be invalid. Hence, all configurations produced by the SAT solver are valid.

\subsection{Performance Evaluation}\label{sect:performance}
\smallskip\noindent\textbf{Experimental Setup.}\quad
All experiments were conducted on a commodity laptop equipped with an Intel Core i5-1135G7@2.40\,GHz CPU and 16\,GB of RAM\@. 
This choice highlights the lightweight nature of our approach: unlike many program analysis and variability-management tools, which often require dedicated servers or high-performance hardware, \ourtool{RustyEx} can be executed on standard developer workstations.
Most of the analyzed projects are organized as \texttt{Cargo} workspaces, which naturally decompose the system into multiple crates. For each crate, we executed \ourtool{RustyEx} with a timeout of \(10\) minutes, which is a realistic bound for integration in a continuous integration (CI) pipeline. Various metrics were collected and then aggregated at the project level, including counts of features and dependencies, statistics on UIR nodes and edges, and summaries of the feature dependency graph and the atom dependency tree.
We also monitored peak memory usage and execution time.
Overall, these choices demonstrate that \ourtool{RustyEx} can be seamlessly adopted in everyday development workflows, where fast turnaround and modest resource requirements are essential.
A replication package with all experimental data and scripts is available on Zenodo: \begin{center}\url{https://doi.org/10.5281/zenodo.17691776}.\end{center}

\begin{figure}[t]
   \centering
   \caption{Correlation between lines of code (x-axis), execution time (s) \protect\tikz\protect\fill[Fig3Blue] (0,0) circle (0.1cm);, and peak memory usage (MB) \protect\tikz\protect\fill[Fig3Orange] (0,0) circle (0.1cm);.}%
   \includegraphics[width=\linewidth, keepaspectratio]{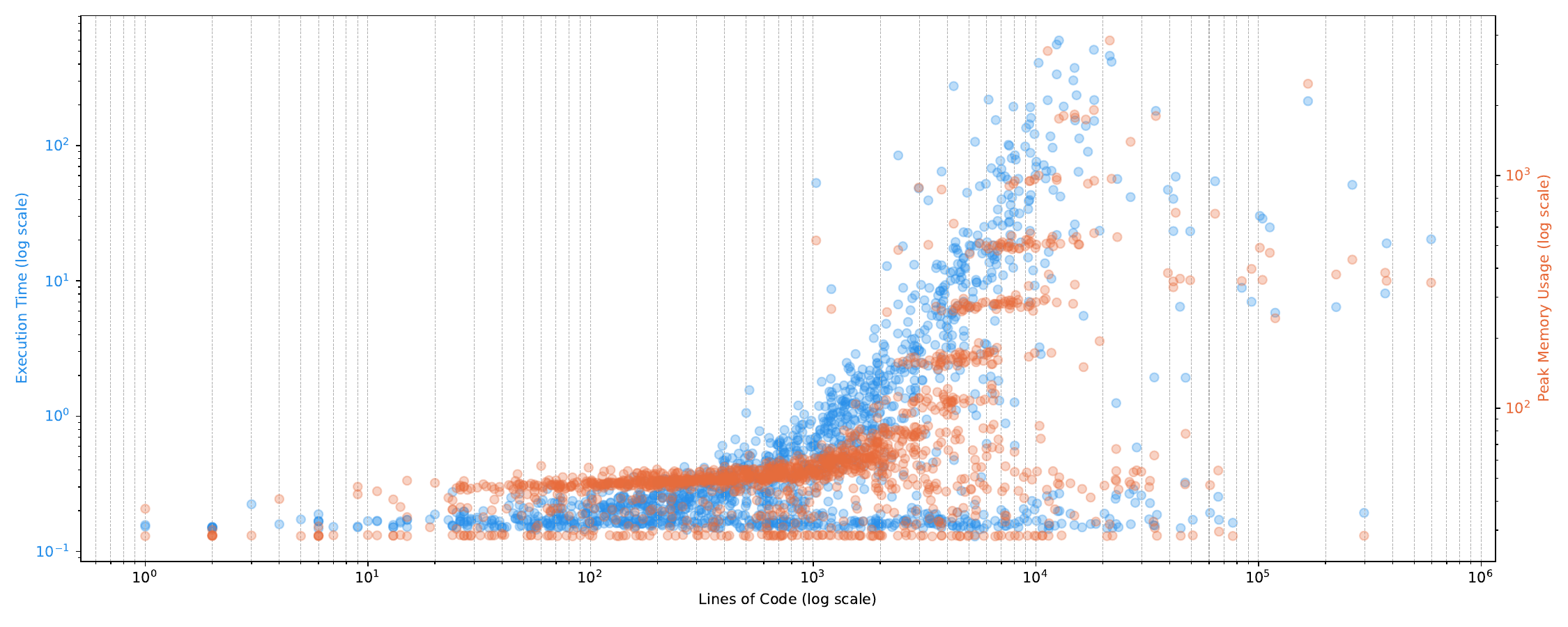}%
   \label{fig:time_loc}
\end{figure}

\smallskip\noindent\textbf{Execution Time and Scalability.}\quad
Figure~\ref{fig:time_loc}---highlighted in~\protect\tikz\protect\fill[Fig3Blue] (0,0) circle (0.1cm);---shows the correlation between lines of code and execution time on a logarithmic scale.
\ourtool{RustyEx} successfully completed the analysis of about \(93\%\) of the projects, showing strong scalability across a wide range of sizes and domains. Execution time scaled smoothly with code size: for small to medium projects, the analysis often completed within seconds, while for the largest ones (e.g., \texttt{hyperswitch} and \texttt{swc}) execution time peaked at \(33\)\,minutes.
The average execution time per project was \(333\)\,seconds, with individual crates requiring about \(8\)\,seconds on average.
These results confirm that our tool is not only efficient in practice, but also robust enough to handle large modular codebases with hundreds of thousands of lines of code and thousands of features.
Importantly, the logarithmic trend visible in the plot suggests that the analysis cost grows sub-linearly compared to project size, which indicates that \ourtool{RustyEx} is suitable for long-term scalability.

\smallskip\noindent\textbf{Memory Usage.}\quad
Figure~\ref{fig:time_loc}---highlighted in \protect\tikz\protect\fill[Fig3Orange] (0,0) circle (0.1cm);---shows the correlation between lines of code and peak memory usage, also on a logarithmic scale.
Memory consumption was generally modest: most projects required less than \(1\)\,GB of RAM, while only a handful of large codebases such as \texttt{hyperswitch} and \texttt{swc} reached a peak of \(3.8\)\,GB\@.
The average peak usage was \(885\)\,MB per project, which fits comfortably within the capacity of mainstream laptops.
At the crate level, average peak memory dropped to \(105\)\,MB, indicating that memory requirements depend more on the internal complexity of each crate than on the overall project size.
This stability across a diverse set of projects confirms that \ourtool{RustyEx} can be executed in environments with limited resources, such as cloud-based CI/CD pipelines or developer laptops.

\smallskip\noindent\textbf{Intermediate Structures.}\quad
The intermediate representations produced by \ourtool{RustyEx} remained compact and efficient to process. On average, the UIR contained approximately \(82{,}000\) nodes and a similar number of edges, which is relatively small compared to the size of the original ASTs.
The \textit{feature dependency graphs} were significantly more compact, typically comprising between \(100\) and \(500\) edges after multi-edge squashing.
Even smaller were the atom dependency trees, which were on average \(95\%\) smaller than the UIR itself.
This compactness is particularly important because it directly impacts the cost of subsequent analyses and ranking procedures: smaller graphs and trees allow for faster computation of centrality measures and logical transformations.
These results highlight that our abstraction strategy, centered on atoms and UIR, strikes a good balance between preserving semantic detail and reducing analysis overhead.

\begin{figure}[t]
   \centering
    \caption{Delta between the number of \textit{declared features} and the number of \textit{detected feature}.}%
   \includegraphics[width=\linewidth, keepaspectratio]{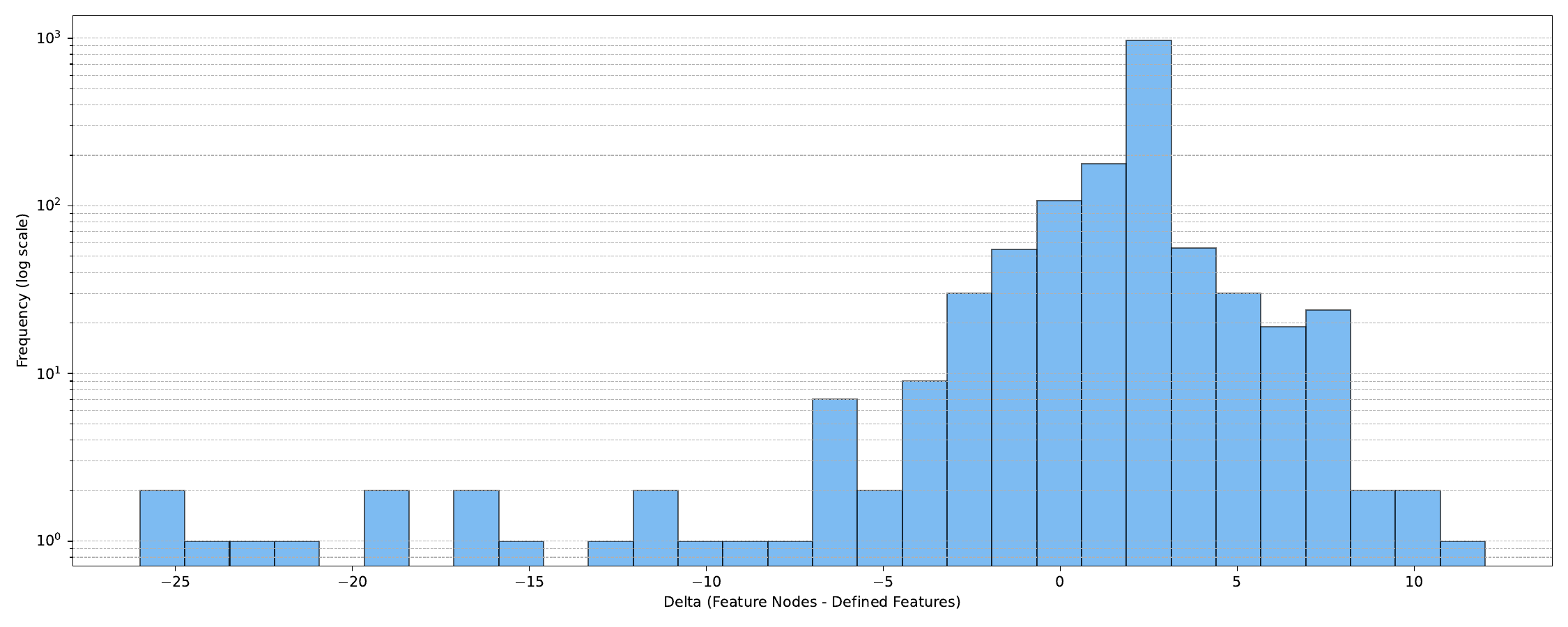}%
   \label{fig:features_delta}
\end{figure}

\smallskip\noindent\textbf{Feature Complexity.}\quad
Most projects defined a substantial number of features, with a median of \(425\) per project. On average, \ourtool{RustyEx} detected \(122\) nodes in the feature dependency graph, often revealing cross-feature dependencies that underscore the challenges of analyzing highly configurable systems.
As shown in Figure~\ref{fig:features_delta}, the number of detected features is generally greater than the number explicitly declared in \texttt{Cargo.toml} files. This discrepancy arises mainly from two factors: \begin{inparaenum} \item feature overlaps across different crates within the same workspace, which are not redefined in the manifest, and \item the handling of the \inlinerust{not} predicate, which introduces logically independent variants and, in the worst case, doubles the number of detected features.\end{inparaenum}

Despite the substantial number of features, their utilization remained modest: on average, only \(228\) feature-related artifacts were identified, which means that each feature was used fewer than two times on average.
This finding suggests that while projects expose a rich configuration space at the feature level, many features have limited practical impact on the actual codebase.
In turn, this motivates the need for principled prioritization: if most features are rarely used, developers and testers should concentrate on the ones that have the greatest influence on the system.

\smallskip\noindent\textbf{Robustness.}\quad
Finally, we assessed the robustness of \ourtool{RustyEx}. Out of more than \(1{,}600\) crates analyzed (roughly \(40\) per project), only \(116\) failed, yielding a \(93\%\) success rate.
Most failures were attributable to missing system dependencies during compilation, which are unrelated to the analysis itself.
Timeouts were rare, affecting fewer than \(2\%\) of the crates.
Importantly, no major crashes or incorrect results were observed. Even large modular workspaces such as \texttt{diem} and \texttt{deno} were analyzed successfully without manual intervention.
These results confirm that \ourtool{RustyEx} is mature and reliable enough to be integrated into the toolchains of both researchers and practitioners.

\section{Threats to Validity}\label{sect:threats-to-validity}
We organize our discussion following Wohlin \textit{et al.}~\cite{Wohlin12}'s taxonomy.

\subsection{Construct Validity}\label{sect:construct-validity}
\noindent\textbf{Proxy metrics for feature relevance.}\quad
We rely on \textit{geometric} and \textit{spectral} centrality computed on the feature dependency graph to approximate feature relevance.
These metrics are not neutral: each has biases and limitations, especially regarding disconnected components and local vs.\ global influence.

\textit{Mitigation.}
\ourtool{RustyEx} allows users to select the centrality measure that best fits their project domain, thus reducing the risk of systematic misinterpretation.
To avoid excluding isolated features, the tool introduces a \textit{patch node} that connects otherwise disconnected components, ensuring that all features are represented in the ranking process.

\smallskip\noindent\textbf{Weighting of UIR nodes.}\quad
In cases where definitions are missing or cycles are detected, nodes of type \textit{reference} are assigned a default weight.
This approximation may underestimate the role of unresolved external calls, potentially distorting prioritization.

\textit{Mitigation.}
\ourtool{RustyEx} implements a recovery mechanism that assigns fallback weights, preventing infinite loops and limiting distortion.
This ensures that unresolved dependencies do not cancel the influence of otherwise relevant nodes.

\subsection{Internal Validity}\label{sect:internal-validity}
\noindent\textbf{Assumptions about SAT solver correctness.}\quad
Our configuration generation relies on the soundness and completeness of the underlying SAT solver.
If the solver produced incorrect results, the validity of generated configurations would be at risk.

\textit{Mitigation.}
To minimize this threat, we use widely adopted and thoroughly tested SAT solvers, reducing the likelihood of errors and strengthening the robustness of our approach.

\smallskip\noindent\textbf{Timeouts and analysis failures.}\quad
A fraction of crates ($\sim7\%$) failed due to timeouts or missing system dependencies.
Such failures may introduce selection bias, as not all crates are equally represented in the analysis.

\textit{Mitigation.}
We adopt a fixed $10$-minute timeout per crate to ensure fairness across the dataset.
Failed crates are skipped, but since the majority of failures were due to missing dependencies rather than intrinsic tool limitations, the overall validity of the evaluation remains preserved.

\subsection{External Validity}\label{sect:external-validity}
\noindent\textbf{Focus on open-source Rust projects.}\quad
Our evaluation is based on $40$ open-source projects from GitHub and \texttt{crates.io}.
While these projects cover a broad spectrum of real-world software, they may not capture the full variability of proprietary or industrial codebases.

\textit{Mitigation.}
Many of the analyzed projects are widely used in production and serve as dependencies for industrial systems.
This increases confidence that the results generalize beyond purely academic or hobbyist software.
Furthermore, since our method is language-agnostic, future work will investigate applications in other ecosystems such as C/C++.

\smallskip\noindent\textbf{Fixed configuration budget.}\quad
We generated a fixed number $K$ of configurations per project.
In practice, real-world scenarios may require adaptive strategies that vary the number of generated configurations depending on project size, release stage, or available resources.

\textit{Mitigation.}
\ourtool{RustyEx} supports parameterized configuration policies, enabling users to adjust $K$ according to their specific needs and constraints.
This flexibility ensures that the tool can be adapted to dynamic development and testing scenarios.

\subsection{Conclusion Validity}\label{sect:conclusion-validity}
\noindent\textbf{Dependence on feature ranking.}\quad
The order in which configurations are generated depends on both the selected centrality metric and the refinements applied to the atom dependency tree.
Different choices may therefore lead to different prioritized sets.

\textit{Mitigation.}
\ourtool{RustyEx} makes it possible to switch among centrality measures and disable refinements.
This allows users to conduct sensitivity analyses and evaluate how rankings vary under different assumptions.

\smallskip\noindent\textbf{No ground truth for configurations.}\quad
There is no universally accepted reference set of ``correct'' or ``important'' configurations.
As a result, it is challenging to directly evaluate the relevance of the configurations produced by our method.

\textit{Mitigation.}
To address this, \ourtool{RustyEx} enforces structural consistency checks and ensures compliance with \texttt{Cargo.toml} constraints, such as required features and valid predicate logic.
Additionally, it verifies that all generated configurations are satisfiable with respect to the derived CNF formula.
These safeguards ensure that generated configurations are both valid and representative of critical execution paths.

\section{Related Work}\label{sect:related-work}
Prioritizing configurations in highly configurable systems has been extensively studied, though from different perspectives and often under different assumptions.
Several comprehensive surveys provide an overview of the field, such as~\cite{Agh24} (see in particular Sect.~4.4),~\cite{ElSharkawy19}, and~\cite{Idham23}, which systematically map research trends and highlight open challenges.
Building on these, we briefly summarize the most relevant contributions along three main dimensions.

\smallskip\noindent\textbf{Static Analysis and Preprocessing.}\quad
Early work on highly configurable systems, especially in the context of the Linux kernel, focused on static analysis and preprocessing techniques.
El Sharkawy \textit{et al.}~\cite{ElSharkawy20} developed methods to process \inlinec{#ifdef} directives for extracting software metrics, making configuration-specific complexity more manageable.
Similarly, Sincero \textit{et al.}~\cite{Sincero10} investigated preprocessing of C macros to detect dead code, thus helping to reduce variability-induced maintenance effort.
These approaches are closely related to our idea of analyzing variability from the source, but they remain tied to metrics extraction or defect detection in specific ecosystems.
By contrast, \ourtool{RustyEx} generalizes the notion of variability-aware static analysis beyond macros or directives, targeting the identification of \emph{relevant configurations} in Rust projects without assuming language-specific preprocessing artifacts.

\smallskip\noindent\textbf{Feature Model-based Prioritization.}\quad
Another major line of research builds on explicit feature models, where features and their relationships are captured in structured diagrams.
Bagheri \textit{et al.}~\cite{Bagheri10} proposed the \textit{stratified analytic hierarchy process} to prioritize and select features by decomposing the decision process into manageable layers.
Peng \textit{et al.}~\cite{Peng16} employed a directed, weighted acyclic graph to assess feature importance via \textit{weighted degree centrality}, highlighting asymmetries in feature influence.
Mannion \textit{et al.}~\cite{Mannion21} explored weighting strategies based on variability types, assuming a graph with explicit sources and sinks.
Beyond prioritization, Bagheri \textit{et al.}~\cite{Bagheri11} also assessed the maintainability of feature models using structural metrics such as cyclomatic complexity~\cite{McCabe76}, network centrality~\cite{Newman10}, and classical software engineering metrics~\cite{Fenton91}.
While these approaches are rigorous and effective, they presuppose the existence of a formal feature model—a strong assumption in practice, since many modern systems (including Rust projects) rely on decentralized, implicit forms of variability encoded directly in build manifests and conditional compilation.
\ourtool{RustyEx} differs in this respect, as it does not require a predefined feature model, but instead reconstructs variability information directly from source and build metadata.

\smallskip\noindent\textbf{Centrality Measures for Prioritization.}\quad
A complementary strand of research applies graph-theoretic concepts, especially centrality measures, to variability management and testing.
Mohammed \textit{et al.}~\cite{Mohammed24} ranked configurations within a single feature model using a variety of centrality metrics, thus identifying configurations with disproportionately high influence.
Levasseur \textit{et al.}~\cite{Levasseur24} used centrality, object-oriented metrics, and machine learning to prioritize unit tests, showing that structural graph properties can guide resource allocation in testing.
Ferreira \textit{et al.}~\cite{Ferreira16} went further by proposing \textit{variational} call graphs~\cite{Ferreira15, Johnson09} enriched with centrality information to pinpoint functions that may become vulnerabilities under certain feature combinations.
While all these works demonstrate the value of centrality in highlighting critical elements, they remain tied either to explicit feature models, to testing strategies, or to specialized tasks such as vulnerability detection.
Our approach builds on the same intuition—that centrality can capture relevance—but applies it to a new abstraction level: the atom dependency tree derived from Rust’s variability constructs.
In doing so, \ourtool{RustyEx} extends the applicability of centrality-based prioritization to ecosystems where feature models are implicit and configuration spaces are both large and sparsely exercised.

\section{Conclusion}\label{sect:conclusion}
In this paper, we presented the first general method for prioritizing configurations in highly configurable software systems via a compiler-based refined ranking of features. Unlike previous approaches, our method does not rely on pre-existing feature models and explicitly accounts for feature dependencies and the extent of code affected by each feature.

The method combines inter-procedural static analysis to extract a \textit{unified intermediate representation} (UIR), construction of the \textit{feature dependency graph} and the \textit{atom dependency tree}, centrality-based ranking, and CNF-based SAT solving. This combination enables the identification and generation of the most relevant configurations while reducing the total number of configurations to consider.

To demonstrate the practicality of our method, we implemented it in \ourtool{RustyEx}, a fully automated tool for Rust software. Extensive evaluation on high-profile open-source Rust projects shows that \ourtool{RustyEx} is scalable, robust, and efficient: it handles large codebases with thousands of features, maintains modest memory and runtime requirements, and achieves a high success rate across crates. The approach is language-agnostic and can be applied to other ecosystems with native variability support, such as C/C++ and Java.

By explicitly prioritizing the most relevant features and configurations, our method—and its RustyEx implementation—supports efficient testing, compiler optimizations, program comprehension, variability management, and regression analysis. Generated configurations are valid, sound, and representative of critical execution paths, providing a principled alternative to stochastic or uniform strategies.

Overall, this work introduces the first general, formally sound, and practical approach to configuration prioritization, improving efficiency and fault detection in highly configurable systems and laying the foundation for future extensions to other languages and domains.

\clearpage
\bibliographystyle{plainurl}
\bibliography{local,strings,metrics,programming,software_engineering,software_architecture,dsl,pl,splc,oolanguages,my_work,grammars,ml+nn,security,roles,learning,cop,testing,dsu,distributed_systems,reflection,aosd,foundations,petri-nets,pattern,logic}

\newcommand\sortnoop[1]{}\newcommand{\splps}{\textsf{SP$\!$\reflectbox{L}L$\!$\reflectbox{SP}}}\newcommand{\starpiler}{\ensuremath{\displaystyle\bigstar}piler}
\begin{thebibliography}{10}

\bibitem{Agh24}
Halimeh Agh, Aidin Azamnour, and Sefan Wagner.
\newblock {Software Product Line Testing: A Systematic Literature Review}.
\newblock {\em {Empirical Software Engineering}}, 29(146), September 2024.

\bibitem{AlHajjaji19}
Mustafa Al-Hajjaji, Thomas Th\"um, Malte Lochau, Jens Meinicke, and Gunter
  Saake.
\newblock {Effective Product-Line Testing Using Similarity-Based Product
  Prioritization}.
\newblock {\em {Software and Systems Modeling}}, 18:499--521, February 2019.

\bibitem{Anderson16}
Brian Anderson, Lars Bergstrom, Manish Goregaokar, Josh Matthews, Keegan
  McAllister, Jack Moffitt, and Simon Sapin.
\newblock {Engineering the Servo Web Browser Engine Using Rust}.
\newblock In Tao Xie and Dongmei Zhang, editors, {\em {Proceedings of the 38th
  International Conference on Software Engineering: Software Engineering in
  Practice (ICSE-SEIP'16)}}, pages 81--89, Austin, TX, USA, May 2016. IEEE.

\bibitem{Apel13b}
Sven Apel, Don Batory, Christian K\"astner, and Gunter Saake.
\newblock {\em {Feature-Oriented Software Product Lines}}.
\newblock Springer, April 2013.

\bibitem{Apel11}
Sven Apel, Hendrik Speidel, Philipp Wendler, Alexander von Rhein, and Dirk
  Beyer.
\newblock {Detection of Feature Interactions Using Feature-Aware Verification}.
\newblock In Corina P\u{a}s\u{a}reanu and John Hosking, editors, {\em
  {Proceedings of the 26th International Conference on Automated Software
  Engineering (ASE'11)}}, pages 372--375, Lawrence, KS, USA, November 2011.
  IEEE.

\bibitem{Bagheri10}
Ebrahim Bagheri, Mohsen Asadi, Dragan Gasevic, and Samaneh Soltani.
\newblock {Stratified Analytic Hierarchy Process: Prioritization and Selection
  of Software Features}.
\newblock In Jan Bosch and Jaejoon Lee, editors, {\em {Proceedings of the 14th
  International Software Product Line Conference (SPLC'10)}}, Lecture Notes on
  Computer Science 6287, pages 300--315, Jeju Island, South Korea, September
  2010. Springer.

\bibitem{Bagheri11}
Ebrahim Bagheri and Dragan Gasevic.
\newblock {Assessing the Maintainability of Software Product Line Feature
  Models Using Strutural Metrics}.
\newblock {\em {Software Quality Journal}}, 19(3):576--612, January 2011.

\bibitem{Bavelas50}
Alex Bavelas.
\newblock {Communication Patterns in Task-Oriented Groups}.
\newblock {\em {The Journal of the Acoustical Society of America}},
  22(6):725--730, November 1950.

\bibitem{Benavides10}
David Benavides, Sergio Segura, and Antonio Ruiz-Cort\'es.
\newblock {Automated Analysis of Feature Models 20 Years Later: A Literature
  Review}.
\newblock {\em {Information Systems}}, 35(6):615--636, September 2010.

\bibitem{Bergstra95}
Jan~A. Bergstra and John~V. Tucker.
\newblock {Equational Specifications, Complete Term Rewriting Systems, and
  Computable and Semicomputable Algebras}.
\newblock {\em {Journal of ACM}}, 42(6):1194--1230, November 1995.

\bibitem{Berman79}
Abraham Berman and Robert~J. Plemmons.
\newblock {\em {Nonnegative Matrices in the Mathematical Sciences}}.
\newblock Academic Press, January 1979.

\bibitem{Biere20}
Armin Biere, Katalin Fazekas, Mathias Fleury, and Maximilian Heisinger.
\newblock {CaDiCaL, Kissat, Paracooba, Plingeling and Treengeling Entering the
  SAT Competition 2020}.
\newblock In Tom\'as\u{s} Balyoi, Nils Froleyks, Marijn J.~H. Heule, Markus
  Iser, Matti J\"arvisalo, and Martin Suda, editors, {\em {Proceedings of the
  SAT Competition 2020}}, pages 50--53, Alghero, Italy, July 2020. University
  of Helsinki.

\bibitem{Bloch23}
Francis Bloch, Matthew~O. Jackson, and Pietro Tebaldi.
\newblock {Centrality Measures in Networks}.
\newblock {\em {Social Choice and Welfare}}, 61(2):413--453, April 2023.

\bibitem{Boldi14}
Paolo Boldi and Sebastiano Vigna.
\newblock {Axioms for Centrality}.
\newblock {\em {Internet Mathematics}}, 10(3--4):222--262, September 2014.

\bibitem{Bonacich72}
Phillip Bonacich.
\newblock {Factoring and Weighting Approaches to Status Scores and Clique
  Identification}.
\newblock {\em {Journal of Mathematical Sociology}}, 2(1):113--120, 1972.

\bibitem{Borgatti05}
Stephen~P. Borgatti.
\newblock {Centrality and Network Flow}.
\newblock {\em {Social Networks}}, 25(1):55--71, January 2005.

\bibitem{Boyapati02}
Chandrasekhar Boyapati, Robert Lee, and Martin Rinard.
\newblock {Ownership Types for Safe Programming: Preventing Data Races And
  Deadlocks}.
\newblock In Satoshi Matsuoka, editor, {\em Proceedings of the 17th Annual ACM
  Conference on Object-Oriented Programming, Systems, Languages, and
  Applications (OOPSLA'02)}, pages 211--230, Seattle, WA, USA, November 2002.
  ACM Press.

\bibitem{Chen22}
Shao-Fu Chen and Yu-Sung Wu.
\newblock {Linux Kernel Module Development with Rust}.
\newblock In Yousra Aafer, Shujun Li, and Yulei Wu, editors, {\em {Proceedings
  of the Conference on Dependable and Secure Computing (DSC'22)}}, pages 1--2,
  Edinburgh, United Kingdom, June 2022. IEEE.

\bibitem{Clarke98}
David~G Clarke, John~M Potter, and James Noble.
\newblock {Ownership types for flexible alias protection}.
\newblock In Craig Chambers, editor, {\em {Proceedings of 13th International
  Conference on Object-Oriented Programming Systems, Languages and Applications
  (OOPSLA'98)}}, pages 48--64, Vancouver, BC, CAnada, October 1998. ACM.

\bibitem{Classen13}
Andreas Classen, Maxime Cordy, Pierre-Yves Schobbens, Patrick Heymans, Axel
  Legay, and Jean-Fran\c{c}ois Raskin.
\newblock {Featured Transition Systems: Foundations for Verifying
  Variability-Intensive Systems and Their Application to LTL Model Checking}.
\newblock {\em IEEE Transactions on Software Engineering}, 39(8):1069--1089,
  August 2013.

\bibitem{Clements01}
Paul Clements and Linda Northrop.
\newblock {\em {Software Product Lines: Practices and Patterns}}.
\newblock {Addison-Wesley}, August 2001.

\bibitem{Cohen07}
Myra~B. Cohen, Matthew~B. Dwyer, and Jiangfan Shi.
\newblock {Interaction Testing of Highly-Configurable Systems in the Presence
  of Constraints}.
\newblock In David~S. Rosenblum and Sebastian~G. Elbaum, editors, {\em
  {Proceedings of the 16th International Symposium on Software Testing and
  Analysis (ISSTA'07)}}, pages 129--139, London, United Kingdom, July 2007.
  ACM.

\bibitem{Cooper22}
Keith~D. Cooper and Linda Torczon.
\newblock {\em {Engineering a Compiler}}.
\newblock Morgan Kaufmann, November 2022.

\bibitem{Das18}
Kousik Das, Sovan Samanta, and Madhumangal Pal.
\newblock {Study on Centrality Measures in Social Networks: A Survey}.
\newblock {\em {Social Network Analysis and Mining}}, 8:13:1--13:11, February
  2018.

\bibitem{Dasgupta99}
Sanjoy Dasgupta.
\newblock {Learning Polytrees}.
\newblock In Kathryn~B. Laskey and Henri Prade, editors, {\em {Proceedings of
  the 15th Conference on Uncertainty in Artificial Intelligence (UAI'99)}},
  pages 134--141, Stockholm, Sweden, July 1999. Morgan Kaufmann Publishers Inc.

\bibitem{Een03}
Niklas E\'en and Niklas S\"orensson.
\newblock {An Extensible SAT-Solver}.
\newblock In Enrico Giunchiglia and Armando Tacchella, editors, {\em
  {Proceedings of the 6th International Conference on Theory and Applications
  of Satisfiability Testing (SAT'03)}}, LNCS 2919, pages 502--518, Santa
  Margherita Ligure, Italy, May 2003. Springer.

\bibitem{ElSharkawy17}
Sascha El-Sharkawy, Adam Krafczyk, and Klaus Schmid.
\newblock {An Empirical Study of Configuration Mismatches in Linux}.
\newblock In Myra Cohen and Mathieu Acher, editors, {\em {Proceedings of the
  21st International Systems and Software Product Line Conference (SPLC'17)}},
  pages 19--28, Sevilla, Spain, September 2017. ACM.

\bibitem{ElSharkawy20}
Sascha El-Sharkawy, Adam Krafczyk, and Klaus Schmid.
\newblock {Fast Static Analyses of Software Product Lines: An Example With More
  Than 42,000 Metrics}.
\newblock In Maxime Cordy and Mathieu Acher, editors, {\em {Proceedings of the
  14th International Working Conference on Variability Modelling of
  Software-Intensive Systems (VaMoS'20)}}, pages 1--9, Magdeburg Germany,
  February 2020. ACM.

\bibitem{ElSharkawy19}
Sascha El-Sharkawy, Nozomi Yamagishi-Eichler, and Klaus Schmid.
\newblock {Metrics for Analyzing Variability and Its Implementation in Software
  Product Lines: A Systematic Literature Review}.
\newblock {\em {Information and Software Technology}}, 106:1--30, February
  2019.

\bibitem{Fenton91}
Norman~E. Fenton.
\newblock {\em {Software Metrics: a Rigorous Approach}}.
\newblock London: Chapman \& Hall, 1991.

\bibitem{Ferreira15}
Gabriel Ferreira, Christian K\"astner, J\"urgen Pfeffer, and Sven Apel.
\newblock {Characterizing Complexity of Highly-Configurable Systems with
  Variational Call Graphs: Analyzing Configuration Options Interactions
  Complexity in Function Calls}.
\newblock In {\em {Proceedings of the Symposium and Bootcamp on the Science of
  Security (HotSoS'15)}}, pages 1--2, Urbana-Champaign, IL, USA, April 2015.
  ACM.

\bibitem{Ferreira16}
Gabriel Ferreira, Momin Malik, Christian K\"astner, J\"urgen Pfeffer, and Sven
  Apel.
\newblock {Do \#ifdefs Influence the Occurrence of Vulnerabilities? An
  Empirical Study of the Linux Kernel}.
\newblock In Rick Rabiser and Bing Xie, editors, {\em {Proceedings of the 20th
  International Systems and Software Product-Line Conference (SPLC'16)}}, pages
  65--73, Beijing, China, September 2016. ACM.

\bibitem{Freeman77}
Linton~C. Freeman.
\newblock {A Set of Measures of Centrality Based on Betweenness}.
\newblock {\em {Sociometry}}, 40(1):35--41, March 1977.

\bibitem{Gamma95}
Erich Gamma, Richard Helm, Ralph Johnson, and John Vlissides.
\newblock {\em {Design Patterns: Elements of Reusable Object-Oriented
  Software}}.
\newblock Professional Computing Series. Addison-Wesley, Reading, Ma, USA,
  1995.

\bibitem{Girard87}
Jean-Yves Girard.
\newblock {Linear Logic}.
\newblock {\em {Theoretical Computer Science}}, 50(1):1--101, 1987.

\bibitem{Girard95}
Jean-Yves Girard, Yves Lafont, and Laurent Regnier.
\newblock {\em {Advances in Linear Logic}}.
\newblock Cambridge University Press, July 1995.

\bibitem{Halin19}
Axel Halin, Alexandre Nuttinck, Mathieu Acher, Xavier Devroey, Gilles Perrouin,
  and Benoit Baudry.
\newblock {Test Them All, Is It Worth It? Assessing Configuration Sampling on
  the JHipster Web Development Stack}.
\newblock {\em {Empirical Software Engineering}}, 24(2):674--717, July 2019.

\bibitem{Howson97}
Colin Howson.
\newblock {\em {Logic with Trees: An Introduction to Symbolic Logic}}.
\newblock Routledge, February 1997.

\bibitem{Idham23}
Muhammad Idham, Shahliza Abd~Halim, Dayang Norhayati~Abang Jawawi, Zalmiyah
  Zakaria, Aldo Erianda, and Nachnoer Arss.
\newblock {Test Case Prioritization for Software Product Line: A Systematic
  Mapping Study}.
\newblock {\em {Journal on Informatics Visualization}}, 7(3-2):2126--2134,
  November 2023.

\bibitem{Johnson09}
Neil~F. Johnson.
\newblock {\em {Simply Complexity: A Clear Guide to Complexity Theory}}.
\newblock Oneworld Publications, October 2009.

\bibitem{Kang90}
Kyo~C. Kang, Sholom~G. Cohen, James~A. Hess, William~E. Novak, and A.~Spencer
  Peterson.
\newblock {Feature-Oriented Domain Analysis (FODA) Feasibility Study}.
\newblock Technical Report CMU/SEI-90-TR-21, {Carnegie Mellon University},
  Pittsburgh, Pennsylvania, USA, November 1990.

\bibitem{Karatas10}
Ahmet~Serkan Karata\c{s}, Halit O\u{g}uzt\"uz\"un, and Ali Do\u{g}ru.
\newblock {Global Constraints on Feature Models}.
\newblock In David Cohen, editor, {\em {Proceedings of the 16th International
  Conference on Principles and Practice of Constraint Programming (CP'10)}},
  LNCS 6308, pages 537--551, St. Andrews, Scotland, September 2010. Springer.

\bibitem{Katz53}
Leo Katz.
\newblock {A New Status Index Derived From Sociometric Analysis}.
\newblock {\em {Psychometrika}}, 18(1):39--43, March 1953.

\bibitem{Kennedy05}
Andrew Kennedy and Claudio~V. Russo.
\newblock {Generalized Algebraic Data Types and Object-Oriented Programming}.
\newblock In Richard~P. Gabriel, editor, {\em {Proceedings of 19th ACM
  International Conference on Object-Oriented Programming Systems, Languages
  and Applications (OOPSLA'05)}}, pages 21--40, San Diego, CA, USA, October
  2005. ACM.

\bibitem{Krueger06}
Charles~W. Krueger.
\newblock {New Methods in Software Product Line Practice}.
\newblock {\em Communications of the ACM}, 49(12):37--40, December 2006.

\bibitem{Landherr10}
Andrea Landherr, Bettina Friedl, and Julia Heidemann.
\newblock {A Critical Review of Centrality Measures in Social Networks}.
\newblock {\em {Business an Informtation Systems Engineering}}, 2(5):371--385,
  October 2010.

\bibitem{Lee19}
Jihyun Lee and Sunmyung Hwang.
\newblock {Combinatorial Test Design Using Design-Time Decisions for
  Variability}.
\newblock {\em {Journal of Software Engineering and Knowledge Engineering}},
  29(8):1141--1158, August 2019.

\bibitem{Lehmann81}
Daniel~J. Lehmann and Michael~B. Smyth.
\newblock {Algebraic Specification of Data Types: A Synthetic Approach}.
\newblock {\em {Journal of Mathematical Systems Theory}}, 14(2):97--139,
  December 1981.

\bibitem{Levasseur24}
Marc-Antoine Levasseur and Mourad Badri.
\newblock {Prioritizing Unit Tests Using Object-Oriented Metrics, Centrality
  Measures, and Machine Learning Algorithms}.
\newblock {\em {Innovations in Systems and Software Engineering}}, pages 1--27,
  February 2024.

\bibitem{Li24}
Hongyu Li, Liwei Guo, Yexuan Yang, Shangguag Wang, and Mengwei Xu.
\newblock {An Empirical Study of Rust-for-Linux: The Success, Dissatisfaction,
  and Compromise}.
\newblock In Saurabh Bagchi and Yiying Zhang, editors, {\em {Proceedings of the
  USENIX Annual Technical Conference (USENIX'24)}}, pages 425--443, Santa
  Clara, CA, USA, July 2024. Curran Associates, Inc.

\bibitem{Li21b}
Zhuohua Li, Jincheng Wang, Mingshen Sun, and John C.~S. Lui.
\newblock {MirChecker: Detecting Bugs in Rust Programs via Static Analysis}.
\newblock In Giovanni Vigna and Elaine Shi, editors, {\em {Proceedings of the
  2021 Conference on Computer and Communications Security (CCS'21)}}, pages
  2183--2196, Virtual, South Korea, November 2021. ACM.

\bibitem{Liebig17}
J\"org Liebig, Sven Apel, Andreas Janker, Florian Garbe, and Sebastian Oster.
\newblock {Handling Static Configurability in Refactoring Engines}.
\newblock {\em {Computer}}, 50(7):44--53, 2017.

\bibitem{Liebig15}
J\"org Liebig, Andreas Janker, Florian Garbe, Sven Apel, and Christian
  Lengauer.
\newblock {Morpheus: Variability-Aware Refactoring in the Wild}.
\newblock In Gerardo Canfora and Sebastian Elbaum, editors, {\em {Proceedings
  of the 37th International Conference on Software Engineering (ICSE'15)}},
  pages 380--391, Florence, Italy, May 2015. IEEE.

\bibitem{Lochau12}
Malte Lochau, Sebastian Oster, Ursula Goltz, and Andy Sch\"urr.
\newblock {Model-Based Pairwise Testing for Feature Interaction Coverage in
  Software Product Line Engineering}.
\newblock {\em {Software Quality Journal}}, 20(3-4):567--604, September 2012.

\bibitem{Magnusson07}
Eva Magnusson and G\"orel Hedin.
\newblock {Circular Reference Attributed Grammars---Their Evaluation and
  Applications}.
\newblock {\em {Science of Computer Programming}}, 68(1):21--37, August 2007.

\bibitem{Mannion21}
Mike Mannion and Hermann Kaindl.
\newblock {Using Binary Strings for Comparing Products From Software-Intensive
  Systems Product Lines}.
\newblock In Ina Schaefer and Maurice~H. ter Beek, editors, {\em {Proceedings
  of the 25th International Software Product Line Conference (SPLC'21)}}, pages
  257--266, Leicester, United Kingdom, September 2021. ACM.

\bibitem{Marchiori00}
Massimo Marchiori and Vito Latora.
\newblock {Harmony in the Small-World}.
\newblock {\em {Physica A: Statistical Mechanics and its Applications}},
  285(3-4):539--546, October 2000.

\bibitem{Martin21}
Hugo Martin, Mathieu Acher, Juliana Alves~Pereira, and Jean-Marc J\'ez\'equel.
\newblock {A Comparison of Performance Specialization Learning for Configurable
  Systems}.
\newblock In Ina Schaefer and Maurice~H. ter Beek, editors, {\em {Proceedings
  of the 25th International Software Product Line Conference (SPLC'21)}}, pages
  46--57, Leicester, United Kingdom, September 2021. ACM.

\bibitem{Matsakis14}
Nicholas~D. Matsakis and Felix~S. Klock.
\newblock {The Rust Language}.
\newblock {\em {ACM SIGAda Letters}}, 34(3):103--104, October 2014.

\bibitem{Mazurkiewicz95}
Antoni Mazurkiewicz.
\newblock {Introduction to Trace Theory}.
\newblock In Volker Diekert and Grzegorz Rozenberg, editors, {\em {The Book of
  Traces}}, chapter~1, pages 3--41. World Scientific Publishing, March 1995.

\bibitem{McCabe76}
Thomas~J. McCabe.
\newblock {A Complexity Measure}.
\newblock {\em IEEE Transactions on Software Engineering}, 2(4):308--320,
  December 1976.

\bibitem{Melo16}
Jean Melo, Elvis Flesborg, Claus Brabrand, and Andrzej W{\k{a}}sowski.
\newblock {A Quantitative Analysis of Variability Warnings in Linux}.
\newblock In Ina Schaefer and Vander Alves, editors, {\em {Proceedings of the
  10th International Workshop on Variability Modelling of Software-Intensive
  Systems (VaMoS'10)}}, pages 3--8, Salvador, Brazil, January 2016. ACM.

\bibitem{Mohammed24}
Fathiya Mohammed, Mike Mannion, Hermann Kaindl, and James Patterson.
\newblock {Evaluating the Relative Importance of Product Line Features using
  Centrality Metrics}.
\newblock In Massimo Mecella and Arend Rensink, editors, {\em {Proceedings of
  the 19th International Conference on Software Technologies (ICSOFT 2024)}},
  pages 469--476, Dijon, France, July 2024. SciTe Press.

\bibitem{Newman01}
Mark E.~J. Newman.
\newblock {Scientific Collaboration Networks. II. Shortest Paths, Weighted
  Networks, and Centrality}.
\newblock {\em {Physical Review E}}, 64(1):016132, June 2001.

\bibitem{Newman10}
Mark E.~J. Newman.
\newblock {\em {Networks: An Introduction}}.
\newblock Oxford University Press, first edition, March 2010.

\bibitem{Odersky92}
Martin Odersky.
\newblock {Observers for Linear Types}.
\newblock In Bernd Krieg-Br\"uckner, editor, {\em {Proceedings of the 4th
  European Symposium on Programming (ESOP'92)}}, LNCS 582, pages 390--407,
  Rennes, France, February 1992. Springer.

\bibitem{VonOheimb99}
David {\sortnoop{Oheimb}}von~Oheimb.
\newblock {Hoare Logic for Mutual Recursion and Local Variables}.
\newblock In C.~Pandu Rangan, Venkatesh Raman, and Ramaswamy Ramanujam,
  editors, {\em {Proceedings of the 19th Conference on Foundations of Software
  Technology and Theoretical Computer Science (FSTTCS'99)}}, LNCS 1738, pages
  168--180, Chennai, India, December 1999. Springer.

\bibitem{Olender90}
Kurt~M. Olender and Leon~J. Osterweil.
\newblock {Cecil: A Sequencing Constraint Language for Automatic Static
  Analysis Generation}.
\newblock {\em IEEE Transactions on Software Engineering}, 16(3):268--280,
  March 1990.

\bibitem{Olender92}
Kurt~M. Olender and Leon~J. Osterweil.
\newblock {Interprocedural Static Analysis of Sequencing Constraints}.
\newblock {\em {Transactions on Software Engineering and Methodology}},
  1(1):21--52, January 1992.

\bibitem{Opsahl10}
Tore Opsahl, Filip Agneessens, and John Skvoretz.
\newblock {Node Centrality in Weighted Networks: Generalizing Degree and
  Shortest Paths}.
\newblock {\em {Social Networks}}, 32(3):245--251, July 2010.

\bibitem{Oster10}
Sebastian Oster, Florian Markert, and Philipp Ritter.
\newblock {Automated Incremental Pairwise Testing of Software Product Lines}.
\newblock In Jan Bosch and Jaejoon Lee, editors, {\em {Proceedings of the 14th
  International Software Product Line Conference (SPLC'10)}}, Lecture Notes on
  Computer Science 6287, pages 196--210, Jeju Island, South Korea, September
  2010. Springer.

\bibitem{Parejo16}
Jos\'e~A. Parejo, Ana~B. S\'anchez, Sergio Segura, Antonio Ruiz-Cort\'es,
  Roberto~E. Lopez-Herrejon, and Alexander Egyed.
\newblock {Multi-Objective Test Case Prioritization in Highly Configurable
  Systems: A Case Study}.
\newblock {\em {Journal of Systems and Software}}, 122:287--310, December 2016.

\bibitem{Patel13}
Sachin Patel, Priya Gupta, and Vipul Shah.
\newblock {Combinatorial Interaction Testing with Multi-Perspective Feature
  Models}.
\newblock In Yang Liu and Pang, editors, {\em {Proceedings of the 6th
  International Conference on Software Testing, Verification and Validation
  Workshops (ICSTW'13)}}, pages 321--330, Luxembourg City, Luxembourg, March
  2013. IEEE.

\bibitem{Peng16}
Zhenlian Peng, Jian Wang, Keqing He, and Hongtao Li.
\newblock {An Approach for Prioritizing Software Features Based on Node
  Centrality in Probability Network}.
\newblock In Georgia Kapitsaki and Eduardo Almeida, editors, {\em {Proceedings
  of the 15th International Conference on Software Reuse (ICSR'15)}}, LNCS
  9679, pages 106--121, Limassol, Cyprus, June 2015. Springer.

\bibitem{Perrouin10}
Gilles Perrouin, Sagar Sen, Jacques Klein, Beno{\^\i}t Baudry, and Yves
  le~Traon.
\newblock {Automated and Scalable T-Wise Test CASE Generation Strategies for
  Software Product Lines}.
\newblock In Ana~Rosa Cavalli and Sudipto Ghosh, editors, {\em {Proceedings of
  the 3rd International Conference on Software Testing, Verification and
  Validation (ICST'10)}}, pages 459--468, Paris, France, April 2010. IEEE.

\bibitem{Pierce02}
Benjamin~C. Pierce.
\newblock {\em {Types and Programming Languages}}.
\newblock MIT Press, February 2002.

\bibitem{Pohl11}
Richard Pohl, Kim Lauenroth, and Klaus Pohl.
\newblock {A Performance Comparison of Contemporary Algorithmic Approaches for
  Automated Analysis Operations on Feature Models}.
\newblock In Corina p\u{a}s\u{a}reanu and John Hosking, editors, {\em
  {Proceedings of the 26th International Conference on Automated Software
  Engineering (ASE'11)}}, pages 313--322, Lawrence, KS, USA, November 2011.
  IEEE.

\bibitem{Prehofer97}
Christian Prehofer.
\newblock {Feature-Oriented Programming: A Fresh Look at Objects}.
\newblock In Mehmet Ak\c{s}it and Satoshi Matsuoka, editors, {\em {Proceedings
  of the 11th European Conference on Object-Oriented Programming (ECOOP'97)}},
  Lecture Notes in Computer Science 1241, pages 419--443, Helsinki, Finland,
  June 1997. Springer.

\bibitem{Qu08}
Xiao Qu, Myra~B. Cohen, and Gregg Rothermel.
\newblock {Configuration-Aware Regression Testing: An Empirical Study of
  Sampling and Prioritization}.
\newblock In Andreas Zeller, editor, {\em {Proceedings of the 17th
  International Symposium on Software Testing and Analysis (ISSTA'08)}}, pages
  75--86, Seattle, WA, USA, July 2008. ACM.

\bibitem{Sanchez17}
Ana~B. S\'anchez, Sergio Segura, Jos\'e~A. Parejo, and Antonio Ruiz-Cort\'es.
\newblock {Variability Testing in the Wild: The Drupal CASE Study}.
\newblock {\em {Software and Systems Modeling}}, 16(2):173--194, April 2017.

\bibitem{Sanchez14}
Ana~B. S\'anchez, Sergio Segura, and Antonio Ruiz-Cort\'es.
\newblock {A Comparison of Test Case Prioritization Criteria for Software
  Product Lines}.
\newblock In Lauric Williams and Claes Wohlin, editors, {\em {Proceedings of
  7th International Conference on Software Testing, Verification and Validation
  (ICST'14)}}, pages 41--50, Cleveland, OH, USA, March 2014. IEEE.

\bibitem{Seeley49}
John~R. Seeley.
\newblock {The Net of Reciprocal Influence; A Problem in Treating Sociometric
  Data}.
\newblock {\em {Canadian Journal of Psychology}}, 3(4):234--240, December 1949.

\bibitem{Seidl16}
Christoph Seidl, Tim Winkelmann, and Ina Schaefer.
\newblock {A Software Product Line of Feature Modeling Notations and Cross-Tree
  Constraint Languages}.
\newblock In Andreas Oberweis and Ralf~H. Reussner, editors, {\em {Proceedings
  of the 13th Edition of Modellierung (Modellierung'16)}}, LNI P-254, pages
  157--172, Karlsruhe, Germany, March 2016. Springer.

\bibitem{Siegmund15}
Norbert Siegmund, Alexander Grebhanhn, Sven Apel, and Christian K\"astner.
\newblock {Performance-Influence Models for Highly Configurable Systems}.
\newblock In Mark Harman and Patrick Heymans, editors, {\em {Proceedings of the
  10th Joint Meeting on Foundations of Software Engineering (ESEC/FSE'15)}},
  pages 284--294, Bergamo, Italy, September 2015. ACM.

\bibitem{Sincero07}
Julio Sincero, Horst Schirmeier, Wolfgang Schr\"oder-Preikschat, and Olaf
  Spinczyk.
\newblock {Is the Linux Kernel a Software Product Line?}
\newblock In Frank van~der Linden and Bj\"orn Lnndell, editors, {\em
  {Proceedings of the 2nd International Workshop on Open Source Software and
  Product Lines (SPLC-OSSPL'07)}}, Kyoto, Japan, September 2007.

\bibitem{Sincero10}
Julio Sincero, Reinhard Tartler, Daniel Lohmann, and Wolfgang
  Schr\"oder-Preikschat.
\newblock {Efficient Extraction and Analysis of Preprocessor-Based
  Variability}.
\newblock In Jaakko J\"arvi, editor, {\em {Proceedings of the 9th International
  Conference on Generative Programming and Component Engineering (GPCE'10)}},
  pages 33--42, Eindhoven, The Netherlands, October 2010. ACM.

\bibitem{Sorensson05}
Niklas S\"orensson and Niklas E\'en.
\newblock {MiniSat v1.13 – A SAT Solver with Conflict-Clause Minimization}.
\newblock In Fahiem Bacchus, editor, {\em {Proceedings of the 8th International
  Conference on Theory and Applications of Satisfiability Testing (SAT'05)}},
  LNCS 3569, pages 1--2, St. Andrews, Scotland,, June 2005. Springer.
\newblock Poster.

\bibitem{Souto18}
Sabrina Souto and Marcelo d'Amorim.
\newblock {Time-space efficient regression testing for configurable systems}.
\newblock {\em {Journal of Systems and Software}}, 137:733--746, March 2018.

\bibitem{Stroustrup94}
Bjarne Stroustrup.
\newblock {\em {The Design and Evolution of C++}}.
\newblock Addison-Wesley, first edition, March 1994.

\bibitem{Tartler14}
Reinhard Tartler, Christian Dietrich, Julio Sincero, Wolfgang
  Schr\"oder-Preikschat, and Daniel Lohmann.
\newblock {Static Analysis of Variability in System Software: The 90,000
  \#ifdefs Issue}.
\newblock In Garth Gibson and Nickolai Zeldovich, editors, {\em {Proceedings of
  the USENIX Annual Technical Conference (USENIX ATC'14)}}, pages 421--432,
  Philadelphia, PA, USA, June 2014. USENIX Association.

\bibitem{Tartler11}
Reinhard Tartler, Daniel Lohmann, Julio Sincero, and Wolfgang
  Schr\"oder-Preikschat.
\newblock {Feature Consistency in Compile-Time-Configurable System Software:
  Facing the Linux 10,000 Feature Problem}.
\newblock In Gernot Heiser, editor, {\em {Proceedings of the 6th Conference on
  Computer Systems (EuroSys'11)}}, pages 47--60, Salzburg, Austria, April 2011.
  ACM.

\bibitem{Treinish22}
Matthew Treinish, Ivan Carvalho, Georgios Tsilimigkounakis, and Nahum S\'a.
\newblock {rustworkx: A High-Performance Graph Library for Python}.
\newblock {\em {Journal of Open Source Software}}, 7(79):3968:1--3968:32,
  November 2022.

\bibitem{Turner85}
David~A. Turner.
\newblock {Miranda: A Non-Strict Functional Language with Polymorphic Types}.
\newblock In Jean-Pierre Jouannaud, editor, {\em {Proceedings of the 1st
  International Conference on Functional Programming Languages and Computer
  Architecture (FPCA'85)}}, LNCS 201, pages 1--16, Nancy, France, September
  1985. Springer.

\bibitem{Velez20}
Miguel Velez, Pooyan Jamshidi, Florian Sattler, Norbert Siegmund, Sven Apel,
  and Christian K{\"a}stner.
\newblock {ConfigCrusher: Towards White-Box Performance Analysis for
  Configurable Systems}.
\newblock {\em {Automated Software Engineering}}, 27:265--300, August 2020.

\bibitem{Velez21}
Miguel Velez, Pooyan Jamshidi, Norbert Siegmund, Sven Apel, and Christian
  K\"astner.
\newblock {White-Box Analysis Over Machine Learning: Modeling Performance of
  Configurable Systems}.
\newblock In {\em {Proceedings of the 43rd International Conference on Software
  Engineering (ICSE'21)}}, pages 1072--1084, Madrid, Spain, May 2021. IEEE.

\bibitem{Velez22}
Miguel Velez, Pooyan Jamshidi, Norbert Siegmund, Sven Apel, and Christian
  K\"astner.
\newblock {On Debugging the Performance of Configurable Software Systems:
  Developer Needs and Tailored Tool Support}.
\newblock In Daniela Damian and Andreas Zeller, editors, {\em {Proceedings of
  the 44th International Conference on Software Engineering (ICSE'22)}}, pages
  1571--1583, Pittsburgh, USA, May 2022. ACM.

\bibitem{Venkatakeerthy20}
S.~Venkatakeerthy, Rohit Aggarwal, Shalini Jain, Maunendra~Sankar Desarkar, and
  Ramakrishna Updrasta.
\newblock {IR2Vec: LLVM IR Based Scalable Program Embeddings}.
\newblock {\em {ACM Transactions on Architecture and Code Optimization}},
  17(4):32:1--32:27, December 2020.

\bibitem{Wadler90}
Philip Wadler.
\newblock {Linear Types Can Change the World!}
\newblock In Manfred Broy and Cliff~B. Jones, editors, {\em {Proceedings of the
  2nd Working Conference on Programming Concepts and Methods (IFIP'90)}}, pages
  561--582, Sea of Galilee, Israel, April 1990. North-Holland.

\bibitem{Wohlin12}
Claes Wohlin, Per Runeson, Martin H\"ost, Magnus~C. Ohlsson, Bj\"orn Regnell,
  and Anders Wessl\'en.
\newblock {\em {Experimentation in Software Engineering}}.
\newblock Springer, 2012.

\end{thebibliography}


\newcommand\sortnoop[1]{}\newcommand{\splps}{\textsf{SP$\!$\reflectbox{L}L$\!$\reflectbox{SP}}}\newcommand{\starpiler}{\ensuremath{\displaystyle\bigstar}piler}

\end{document}